\documentclass[11pt,twoside]{article}

\usepackage{epsfig}
\usepackage{fancyhdr}
\usepackage{latexsym}
\usepackage{amsmath}
\usepackage{amssymb}
\usepackage[latin1]{inputenc}
\usepackage{layout}
\newcommand{\qqbar}{\ensuremath{\rm q\bar{q}}}

\newcommand{\sq}{\ensuremath{{\tilde{\rm q}}}}
\newcommand{\glu}{\ensuremath{{\tilde{\rm g}}}}

\newcommand{\sqqbar}{\ensuremath{\sq\bar{\rm q}}}

\def\Journal#1#2#3#4{{#1} {\bf #2}, #3 (#4)}
\def\etal{et al.}

\def\NPB{{\em Nucl. Phys.} {\bf  B}}

\def\PLB{{\em Phys. Lett.} {\bf  B}}

\def\PRL{\em Phys. Rev. Lett.}

\def\PRD{{\em Phys. Rev.} {\bf D}}

\def\EPJ{{\em Eur. Phys. J.} {\bf C}}
\def\PREP{\em Phys. Rept.}

\def\RNC{\em Riv. Nuovo Cim.}
\def\JMA{{\em Int. J. Mod. Phys.} {\bf A}}

\begin{document}

\pagenumbering{arabic}
\pagestyle{plain}
\Large\hspace*{10.95cm}\parbox{5cm}{hep-ex/0404001\\April 2004}

\vspace*{3cm}

\begin{center}
\LARGE{Interactions of Heavy Stable Hadronizing Particles}
\vspace*{.3cm}

\normalsize A.C.\,Kraan 

\emph{The Niels Bohr Institute, University of Copenhagen, \\Blegdamsvej 17, 2100 Copenhagen-E, Denmark\\ email: ackraan@nbi.dk}

\vspace*{.3cm}
\end{center}



\vspace{.2cm}
\begin{abstract}
In this article, we study the interactions of stable, hadronizing new states, arising in certain extensions of the Standard Model. A simple model, originally intended for stable gluino hadrons, is developed to describe the nuclear interactions of hadrons containing any new colour triplet or octet stable parton. Hadron mass spectra, nuclear scattering cross sections and interaction processes are discussed. Furthermore, an implementation of the interactions of heavy hadrons in GEANT 3 is presented, signatures are studied, and a few remarks about possible detection with the ATLAS experiment are given.
\end{abstract}

\vfill
\normalsize
\newpage
\section{Introduction}
Among the more plausible scenarios of physics beyond the Standard Model is supersymmetry (SUSY)~\cite{susy}. In conventional SUSY models, the lightest supersymmetric particle (LSP) is neutral and colourless. However, models exist in which the LSP is coloured. Models with a gluino LSP are reviewed in Ref.~\cite{gunion}, and include gauge mediated supersymmetry breaking models and string motivated supersymmetric models. Stable squarks are cosmologically disfavoured because of their nonzero electric  charge~\cite{ellis}, but could be sufficiently long-lived to behave like an effectively stable LSP in a detector experiment. A coloured LSP would hadronize into heavy (charged and neutral) bound states. These bound states (for example \glu g, \glu \qqbar, \glu qqq, \sqqbar, \sq qq) are generically called R--hadrons, where the ``R'' refers to the fact that they carry one unit of R--parity~\cite{Farrar:1978xj}. In this article, we shall focus on the fact that such a hadronized LSP will have measurable interactions in a detector, in contrast to conventional SUSY studies, in which the LSP is a weakly-interacting particle, i.e.~a neutralino or a sneutrino. 

\noindent In addition to supersymmetry, other extensions of the Standard Model have been proposed, which predict the existence of new heavy hadrons, either due to the presence of a new conserved quantum number, or because the decays are suppressed by kinematics or couplings. For example, theories with leptoquarks predict stable hadronized states, if the Yukawa coupling between the leptoquark and its decay products is so small that the leptoquark is long-lived~\cite{Leptoquarks}. Another example is theories with universal extra dimensions, where exact momentum conservation in all dimensions leads to stable Kaluza-Klein excitations of e.g.~quarks and gluons, which would form stable hadronizing states~\cite{Appelquist:2000nn}. 

\noindent Searches for stable massive particles are summarized e.g. in Ref.~\cite{massive}, which includes different search methods. Negative results of searches in ordinary matter and cosmic rays imply that the relic density of such particles must be too small for detection. In Ref.~\cite{gunion} it is argued that an LSP gluino can indeed have a very small relic density. Alternatively, the cosmological bounds on stable massive particles can be evaded if the hadron decays slowly. In this paper, we refer to particles as \emph{stable} as long as they are stable from a detector point of view. Concerning search methods using accelerators, these are limited by the center of mass energy of the accelerator and are concentrated on charged heavy particle searches~\cite{charged}. A few searches for neutral hadrons, in particular R--hadrons, exist~\cite{rhadrons1, rhadrons2}. From accelerator searches it is clear that, if heavy hadrons with masses less than the order of a hundred GeV had existed, they would have been detected. In this paper, particles in which we are interested, and which we classify as \emph{heavy}, are particles with masses $\gtrsim$ 100 GeV. 

\noindent The energies of heavy hadrons produced at a $\textrm{p}-\textrm{p}$ collider like the LHC are typically about two to three times their own mass, i.e.~they may be relativistic but their mass is still far from negligible. On their trajectory through the detector, they slow down even further.

\noindent Although electromagnetic interactions of heavy states are well understood, little information is available about nuclear interactions of heavy hadrons, and the resulting energy deposits in a detector calorimeter. Models to describe the nuclear interactions of gluino bound states in particular have previously been discussed in Refs. ~\cite{gunion, Mafi:1999dg, Raby:1998xr}. Here, a new model for the interactions of gluino hadrons is presented, which is more generally applicable to any kind of heavy hadrons, and which provides a convenient basis for future refinements.

\noindent The organization of this paper is as follows. First, electromagnetic interactions of charged heavy hadrons will be discussed in Section~\ref{sec:ref3}. Next, in Section 3, we propose a model describing nuclear interactions of heavy hadrons. Aspects such as hadron mass spectra and  nuclear scattering cross-sections will be discussed. Section 4 is devoted to the GEANT 3 simulation of R--hadronic nuclear interactions. In Section~\ref{sec:pros}, results of the simulation will be presented, together with studies of typical detector signals. In Section~\ref{atlasref}, we briefly discuss manifestations of R--hadrons in the ATLAS experiment, and finally in Section~\ref{con}, we summarize and discuss future studies. 
\section{Electromagnetic interactions of heavy charged hadrons}\label{sec:ref3}A charged heavy particle suffers both continuous ionization losses as well as repeated Coulomb scatterings~\cite{pdg}. Continuous ionization losses of heavy particles are known to be described by the Bethe-Bloch equation. Since electromagnetic losses are proportional to 1/$\beta^2$, the losses for a particle moving with $\beta\ll$1 are considerable. The ultra-relativistic rise in electromagnetic energy losses is not relevant here, as can be seen from Fig~\ref{fig:1m}. Repeated Coulomb scatterings change the particle trajectory. The deflection angle, being proportional to 1/$\beta p$, is small, the small velocity being compensated for by the large momentum with which a heavy particle is normally produced.
\section{Nuclear interactions of heavy hadrons}\label{sec:ref4}
Both charged and neutral heavy hadrons lose energy through scattering off nuclei. In the following, a simple and general framework is presented for simulating nuclear interactions of heavy  hadrons, independent of the new physics model in which the hadron arises. Exotic colour triplet states will be denoted by $\textrm{C}_3$, colour antitriplet states by $\textrm{C}_{\bar{3}}$ and colour octet states by $\textrm{C}_8$. When referring to the mass of a parton, we always mean the constituent mass. An interpretation in terms of a pole mass or other mass definitions is clearly irrelevant for the qualitative considerations in this paper. For simplicity, we take into account only states containing $\textrm{u}$ and $\textrm{d}$ quarks, with constituent masses assumed to be $m_\textrm{q}=0.3$ GeV.
\subsection{The role of the heavy object}\label{sec:ref} 
\begin{figure}[t!]
\begin{center}
\begin{tabular}{ccc}
\epsfig{file=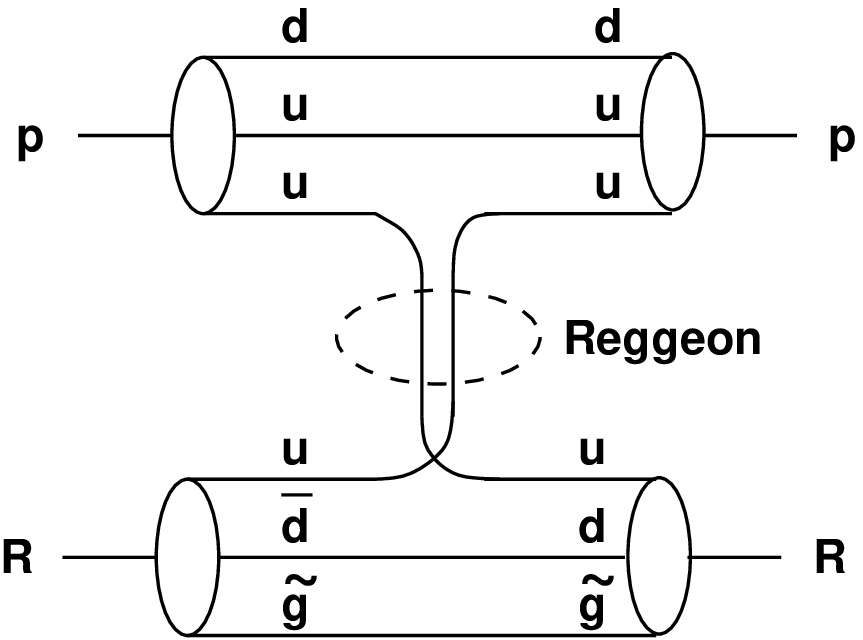,height=3.2cm,width=4.3cm}&\hspace{0.4cm}\epsfig{file=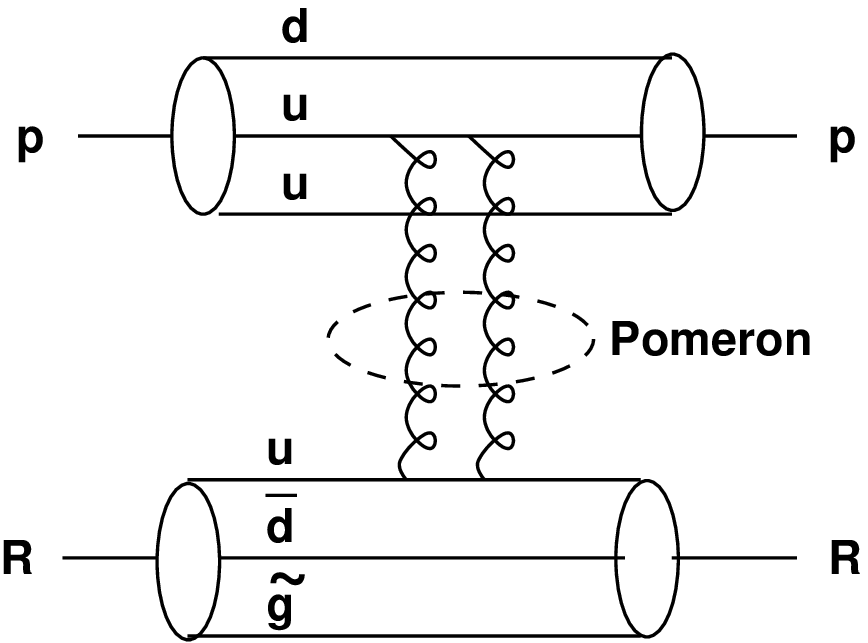,height=3.2cm,width=4.3cm} &\hspace{0.4cm}\epsfig{file=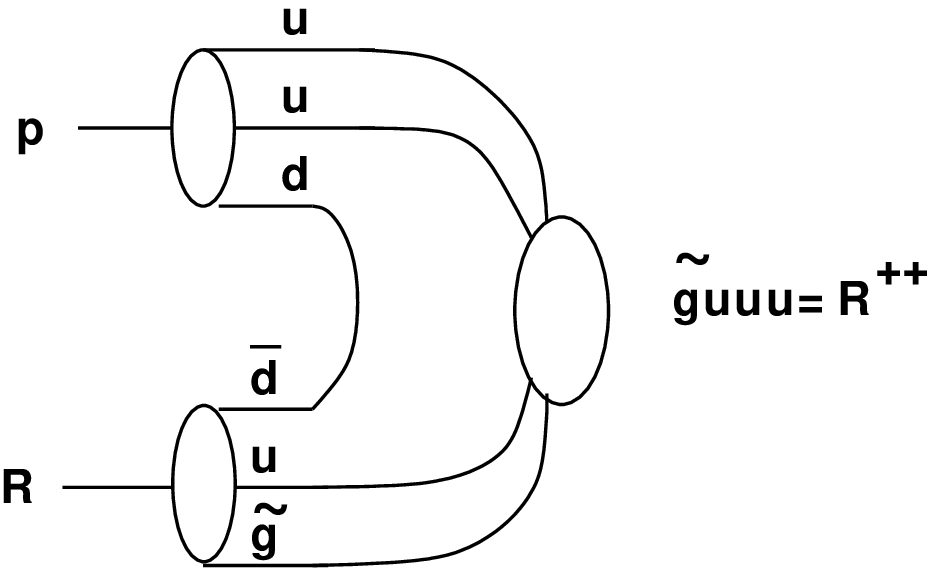,height=3.2cm,width=4.3cm}\\
(a) & (b) & (c)
\end{tabular}
\end{center}
\caption{R--hadron scattering of a proton. (a) Reggeon mediated elastic scattering, (b) Pomeron mediated elastic scattering, (c) the formation of nuclear resonances \label{fig:processesr}}
\end{figure}

Any stable heavy coloured exotic particle hadronizes and forms a colour singlet, for example $\textrm{C}_3\bar{\textrm{q}}$, $\textrm{C}_3\textrm{q}\textrm{q}$, $\textrm{C}_{\bar{3}}\textrm{q}$, $\textrm{C}_8\textrm{q}\bar{\textrm{q}}$, $\textrm{C}_8\textrm{q}\textrm{q}\textrm{q}$, $\textrm{C}_8\textrm{g}$, etc. The probability that the parton $\textrm{C}_i$ of colour state $i$ will interact perturbatively with the quarks in the target nucleon is small, since such interactions are suppressed by the squared inverse mass of the parton. As a consequence, the heavy hadron can be seen as consisting of an essentially non-interacting heavy state $\textrm{C}_i$, accompanied by a coloured hadronic cloud of light constituents, responsible for the interaction. The effective interaction energy of the hadron is therefore small, as can be seen by considering a $\textrm{C}_8\textrm{q}\bar{\textrm{q}}$ state, e.g.~with a total energy $E$=450 GeV and a mass $m$ of the  $\textrm{C}_8$ parton of 300 GeV, i.e.~with a Lorentz factor of $\gamma$=1.5. Although the kinetic energy of the hadron is 150 GeV, the kinetic energy of the interacting $\textrm{q}\bar{\textrm{q}}$ system is only $(\gamma-1) m_{\textrm{q}\bar{\textrm{q}}}\approx$ 0.3 GeV, (if the quark system consists of up and down quarks). Thus, the energy scales relevant for heavy hadron scattering processes off nucleons are low! It is therefore most likely that interaction processes are mediated by Reggeons (Fig.~\ref{fig:processesr}a) and not by by Pomerons (Fig.~\ref{fig:processesr}b). In conclusion, it becomes apparent that the presence of the heavy parton $\textrm{C}_i$ has two basic consequences.
\begin{itemize}
\item It acts as a reservoir of kinetic energy. After an interaction, where the light interacting system loses kinetic energy, new kinetic energy is transferred to it from the co-moving $\textrm{C}_i$-parton.
\item{The parton $\textrm{C}_i$ forces the quark system to be in a certain colour-state, in order for the system as a whole to form a colour-singlet state. This influences the mass spectrum of the hadrons, as will be explored in detail in Section~\ref{hadrons}.  }
\end{itemize}
The following issues are not influenced by the presence of the heavy parton: 
\begin{itemize}\item The interactions of the light constituents are of the same character as for an ordinary hadron. As stressed in the beginning of this section, the heavy state $\textrm{C}_i$ acts as spectator in an interaction. \item The overall size of the hadron as a whole is probably not strongly influenced. This can be understood by recalling that the wave-function associated with a particle with mass $M$ roughly scales as $1/M^2$. The influence of the heavy object $\textrm{C}_i$ on the size of the total hadron is therefore minimal, and the light quarks dominate completely. 
\end{itemize}
\subsection{Hadron mass spectrum} \label{hadrons}
In order to establish the signatures of heavy hadrons in a detector, the mass spectrum of the states must be understood. For example, if the C$_8\textrm{uuu}$ state were to be the lightest baryonic state, then decays into it would be kinematically favourable. This would be important, since the electromagnetic energy losses would then be increased significantly (if the $\textrm{C}_8$ parton does not carry negative charge). A completely different signature would result if a neutral C$_8\textrm{udd}$ state would be the lightest. 

\noindent The mass of the lowest-lying hadronic states, with no radial excitation or orbital angular momentum, can be approximated by (Ref.~\cite{Close})
\begin{equation}\label{eq:hadron}
\hspace{-1.8cm}m_{\textrm{hadron}}=\sum_im_i - k\sum_{i\neq j}\frac{(\mathbf{F_i}\cdot\mathbf{F_j})(\mathbf{S_i}\cdot\mathbf{S_j})}{m_i m_j}
\end{equation}
in which the summation is over all partons $i$ contained in the hadron, $m_i$ is the parton constituent mass, $\mathbf{F}_i$ is the SU(3) colour matrix for parton $i$, $\mathbf{S}_i$ is the SU(2) spin matrix for parton $i$, and $k$ is a constant with dimension (mass)$^3$. The second term, responsible for the mass splitting, depends on the colour-state and spin-state of the hadron. In the derivation of the hadron mass spectrum of heavy hadrons, terms involving the heavy parton in the denominator can be neglected. Expressions for $(\mathbf{F}_1\cdot \mathbf{F}_2)$ and for $(\mathbf{S}_1\cdot \mathbf{S}_2)$ can easily be derived for a $\textrm{qq}$ and $\textrm{q}\bar{\textrm{q}}$ state~\cite{Close}. A $\textrm{qq}$ state can either form a colour antitriplet or a sextet configuration ($\mathbf{3}\otimes\mathbf{3}=\mathbf{\bar{3}}\oplus\mathbf{6}$), while a $\textrm{q}\bar{\textrm{q}}$ state can form a colour singlet or an octet state ($\mathbf{3}\otimes\mathbf{\bar{3}}=\mathbf{1}\oplus\mathbf{8}$). The eigenvalues of the operator $(\mathbf{F}_1\cdot \mathbf{F}_2)$ are $-4/3$, $-2/3$, $1/3$ and $1/6$ when the state is in a colour singlet, antitriplet, sextet and octet configuration, respectively. The eigenvalues of $(\mathbf{S}_1\cdot \mathbf{S}_2)$ are $-3/4$ and $+1/4$ for a $\textrm{qq}$ or $\textrm{q}\bar{\textrm{q}}$ state with total spin zero and spin one, respectively. Eq.~\ref{eq:hadron} leads to the familiar mass patterns of the known hadrons. For example, the $\pi-\rho$, $K-K^*$, $D-D^*$ and $B-B^*$ systems follow Eq.~\ref{eq:hadron} reasonably well, with $k\approx 0.043$ GeV$^3$. 
\noindent Below, the mass spectra of heavy meson and baryon states will be discussed using Eq.~\ref{eq:hadron}. With more refined methods, it would be possible to calculate the spectrum of all possible heavy mesons and baryons precisely. Here, we are mainly interested in the order of magnitude of the mass splittings and not in fine details. 
\subsubsection{Meson mass spectrum}
Possible heavy mesons are $\textrm{C}_{\bar{3}}\textrm{\textrm{q}}$, $\textrm{C}_3\bar{\textrm{\textrm{q}}}$ and $\textrm{C}_8\textrm{\textrm{q}}\bar{\textrm{\textrm{q}}}$ states. For $\textrm{C}_{\bar{3}}\textrm{\textrm{q}}$ and $\textrm{C}_3\bar{\textrm{\textrm{q}}}$ states, no significant mass splitting is expected to occur, since the second term of Eq.~\ref{eq:hadron} is negligible (certainly significantly smaller than that of the $B-B^*$ system, which is only 46 MeV). 

\noindent For a $\textrm{C}_8 \textrm{\textrm{q}}\bar{\textrm{\textrm{q}}}$ state, the mass spectrum is given by:
\begin{eqnarray*}
M_{\textrm{C}_8\textrm{q}\bar{\textrm{q}}}& \approx & M_{\textrm{C}_8}+0.3+0.3-0.043\times\frac{(\frac{1}{6}\times-\frac{3}{4})}{0.3\times 0.3}\hspace{2cm} \textrm{s$_{\textrm{q}\bar{\textrm{q}}}$=0}\\
&\approx& M_{\textrm{C}_8}+0.66\\
M_{\textrm{C}_8\textrm{q}\bar{\textrm{q}}}& \approx& M_{\textrm{C}_8}+0.3+0.3-0.043\times\frac{(\frac{1}{6}\times+\frac{1}{4})}{0.3\times 0.3}\hspace{2cm} \textrm{s$_{\textrm{q}\bar{\textrm{q}}}$=1}\\
& \approx &  M_{\textrm{C}_8}+0.58
\end{eqnarray*}
There are two noticeable aspects. First, the mass hierarchy is reversed, as compared to that of the $\pi-\rho$ mass splitting: the spin-zero state is the heaviest. Second, contrary to the $\pi-\rho$ case, the mass splitting between mesons with different spin is much smaller. Consequently, mass splittings for heavy mesons may safely be neglected. 
\subsubsection{Baryon mass spectrum}
Possible heavy baryons are  $\textrm{C}_3\textrm{q}\textrm{q}$ and $\textrm{C}_{\bar{3}}\bar{\textrm{q}}\bar{\textrm{q}}$, $\textrm{C}_8\textrm{q}\textrm{q}\textrm{q}$ and $\textrm{C}_8\bar{\textrm{q}}\bar{\textrm{q}}\bar{\textrm{q}}$ states. The mass spectra are obtained by a similar calculation as for the mesonic case, with $k\approx0.026$ GeV$^3$, as derived from the ordinary baryon sector. For $\textrm{C}_3\textrm{q}\textrm{q}$ states (and, by symmetry for the $\textrm{C}_{\bar{3}}\bar{\textrm{q}}\bar{\textrm{q}}$ baryons) we obtain the mass spectrum
\begin{eqnarray*}
M_{\textrm{C}_3\textrm{q}\textrm{q}}& \approx & M_{\textrm{C}_3}+0.3+0.3-0.026\times\frac{(-\frac{2}{3}\times-\frac{3}{4})}{0.3\times 0.3}\hspace{2cm} \textrm{s$_{\textrm{q}\textrm{q}}$=0}\\
&\approx& M_{\textrm{C}_3}+0.46\\
M_{\textrm{C}_3\textrm{q}\textrm{q}}& \approx& M_{\textrm{C}_3}+0.3+0.3-0.026\times\frac{(-\frac{2}{3}\times+\frac{1}{4})}{0.3\times 0.3}\hspace{2cm} \textrm{s$_{\textrm{q}\textrm{q}}$=1}\\
& \approx &  M_{\textrm{C}_3}+0.65
\end{eqnarray*}
At this stage we must recall that the total wavefunction associated with a quark system can be decomposed in $flavour\times spin \times colour$, which has to be anti-symmetric for the $\textrm{qq}$ system here. Taking into account that the colour wavefunction, associated with the two quarks, is antisymmetric (it is in an antitriplet configuration), it is seen that the two quarks in a $\textrm{C}_3\textrm{u}\textrm{u}$ or $\textrm{C}_3\textrm{d}\textrm{d}$ state can only be in a symmetric spin-configuration, i.e.~$s_{\textrm{qq}}=1$. This implies that it will kinematically be favourable for the other baryons to decay into the lighter $\textrm{C}_3\textrm{u}\textrm{d}$ state with $s_{qq}=0$ rather than in the heavier $\textrm{C}_3\textrm{u}\textrm{u}$, $\textrm{C}_3\textrm{u}\textrm{d}$ or $\textrm{C}_3\textrm{d}\textrm{d}$ states with $s_{qq}=1$.

\noindent Deriving the mass spectrum of $\textrm{C}_8\textrm{q}\textrm{q}\textrm{q}$ is considerably more complicated. Fortunately, however, we expect the mass splitting to be small. This is based on the following line of arguments.  

\noindent Recall that the total wavefunction associated with the three quarks in a $\textrm{C}_8\textrm{q}\textrm{q}\textrm{q}$ state must be anti-symmetric. For the $\textrm{C}_8\textrm{u}\textrm{u}\textrm{u}$ or $\textrm{C}_8\textrm{d}\textrm{d}\textrm{d}$ states, the three quarks are in a symmetric flavour configuration, and thus the $spin \times colour$ wavefunction should be anti-symmetric. The three quarks being in a colour octet configuration implies that $s_{\textrm{qqq}}$=1/2~\cite{Close}. Thus, the $\textrm{qq}$ states involved in Eq.~\ref{eq:hadron} have spin $s_{\textrm{qq}}=1$ or $s_{\textrm{qq}}=0$ and the associated wavefunctions are mixed symmetric and anti-symmetric. The colour configuration of the $\textrm{qq}$ states is also mixed: either colour antitriplet or sextet. 

\noindent For the $\textrm{C}_8\textrm{u}\textrm{u}\textrm{d}$ and $\textrm{C}_8\textrm{u}\textrm{d}\textrm{d}$ states, the situation is slightly different. The wavefunction associated with flavour can either be totally symmetric or mixed symmetric and anti-symmetric. In the former case, the situation is exactly the same as for the $\textrm{C}_8\textrm{u}\textrm{u}\textrm{u}$ and $\textrm{C}_8\textrm{d}\textrm{d}\textrm{d}$ states. In the latter case, the $spin \times colour$ wavefunction associated to the three quarks must also be mixed. The three quarks being in a colour-octet configuration, such mixed states can be obtained with $s_{\textrm{qqq}}$=1/2 or $s_{\textrm{qqq}}=3/2$~\cite{Close}. The spin doublet case, $s_{\textrm{qqq}}$=1/2, implies a mixed colour configuration and a mixed spin configuration for the three quarks together. For the quark pairs involved in Eq.~\ref{eq:hadron}, this means that they have a mixed $s_{\textrm{qq}}=1$ or $s_{\textrm{qq}}=0$ configuration, and a mixed colour antitriplet and sextet configuration. On the other hand, the spin quartet case, $s_{\textrm{qqq}}=3/2$, implies a mixed colour configuration and a symmetric spin configuration for the three quarks. For the quark pairs involved in Eq.~\ref{eq:hadron}, this means that they have a symmetric $s_{\textrm{qq}}=1$ spin configuration, and again their colour configuration is a mixture of colour antitriplet and sextet.

\noindent Summarizing, $\textrm{C}_8\textrm{u}\textrm{u}\textrm{u}$ or $\textrm{C}_8\textrm{d}\textrm{d}\textrm{d}$ states only arise if $s_{\textrm{qqq}}$=1/2, while $\textrm{C}_8\textrm{u}\textrm{u}\textrm{d}$ and $\textrm{C}_8\textrm{u}\textrm{d}\textrm{d}$ states arise if $s_{\textrm{qqq}}$=1/2 or 3/2. (Note that this is the opposite for ordinary baryons!) 
We saw above that both in the $s_{\textrm{qqq}}=1/2$ case and in the $s_{\textrm{qqq}}=3/2$ case the wavefunctions of the $\textrm{qq}$ combinations involved in Eq.~\ref{eq:hadron} are always strongly mixed: the colour wavefunction is mixed, and either the spin, or the flavour wavefunctions, or both, are mixed for the $\textrm{qq}$ combinations. The colour antitriplet and sextet contributions enter with opposite sign, as do the $s_{\textrm{qq}}=1$ and $s_{\textrm{qq}}=0$ terms. Without having done the exact calculation, it is therefore certain that this feature will result in a  partial cancellation of the mass splitting expression. Thus, the splitting is not expected to be more than the order of a hundred MeV. Exactly the same applies for  $\textrm{C}_8\bar{\textrm{q}}\bar{\textrm{q}}\bar{\textrm{q}}$ states. In conclusion, the mass splittings between the different  $\textrm{C}_8\textrm{q}\textrm{q}\textrm{q}$ and $\textrm{C}_8\bar{\textrm{q}}\bar{\textrm{q}}\bar{\textrm{q}}$ spin states are not expected to be large and may therefore be neglected to first approximation.
\subsection{Resonances}
The formation of resonances is closely connected with the discussion above. In the presence of a colour state $\textrm{C}_i$, new resonances may arise, as illustrated in Fig.~\ref{fig:processesr}c where $\textrm{C}_i=\textrm{C}_8$ is a gluino. We do not have any knowledge about the detailed mechanism of resonance formation in heavy hadrons, but a few remarks can be made. Concentrating on $\textrm{C}_8$ hadrons, if indeed mass splittings between the different hadrons are not larger than the order of a hundred MeV, resonances will not play a significant role. The minimum center-of-mass energy for a scattering of for example a $\textrm{C}_8\textrm{q}\bar{\textrm{q}}$ state with a nucleon with mass $m_n$ is $m_{\textrm{C}_8\textrm{q}\bar{\textrm{q}}}+m_n\approx m_{\textrm{C}_8}+5m_\textrm{d}\approx m_{\textrm{C}_8}+1.5$ GeV, an energy which is above the masses of the most significant resonances. Thus, even if resonances exist, they are not likely to play an important role, but they would be smoothly merged with the continuum, as for $\pi-p$ scattering above the $\Delta$ resonance region. Because of this, and lacking a detailed description of these resonances, we will not take them into account explicitly.


\subsection{Black disk approximation}
Predicting the total cross section of a heavy hadron scattering off a nucleon is non-trivial. Here, a series of arguments will be presented, that will be used as guidelines in constructing an effective model. At high center-of-mass energies, cross sections may be approximated by the geometrical cross section. At low energy, the cross section is hard to estimate; in fact, even for normal hadrons, low energy scattering cross sections are poorly understood.  A vanishing cross section at low energy, as exhibited by  pion-nucleon scattering, is not probable. The reason why this cross section vanishes is that s-wave scattering is forbidden, a unique feature for pion-nucleon scattering~\cite{jackson}. In a $\textrm{C}_8\textrm{q}\bar{\textrm{q}}$ hadron, even though containing $\textrm{u}$ and $\textrm{d}$ quarks, the quark system is in a colour-octet state, and is thus no ordinary pion. In particular, a colour octet pion is not light, as shown in the previous section, and can consequently not be treated as a Goldstone boson; hence s-wave scattering should be possible. On the other hand, a rise in the cross section in for example the nuclear scattering of a $\textrm{C}_8\textrm{q}\textrm{q}\textrm{q}$ state at low energies, as is the case for proton-neutron scattering, is likewise also improbable. The reason why this cross section is so large is again a unique one and is connected to the existence of the deuteron~\cite{jackson} . Expecting neither a rise, nor a vanishing behaviour, the pragmatic solution is to treat a heavy hadron simply as a black disk and to use the geometrical cross section at all scattering center-of-mass energies. 

\subsection{The size of the total cross sections}\label{sec:cross}
The size of the heavy hadron being roughly the same as the size of the accompanying hadron system, the total cross section for nucleon scattering can be approximated by the asymptotic values for the cross sections for normal hadrons scattering off nucleons. For example, in the case of a $\textrm{C}_8\textrm{q}\bar{\textrm{q}}$ state, in which the $\textrm{q}\bar{\textrm{q}}$ system contains only $\textrm{u}$ and $\textrm{d}$ quarks, one can use the asymptotic value of the total cross section for pion-nucleon scattering, while in case one $\textrm{s}$ quark is present, one can use asymptotic values for kaon-nucleon scattering, etc. A simple rule would be 12 mbarn for every $\textrm{u}$ or $\textrm{d}$ quark, and 6 mbarn for every $\textrm{s}$ quark scattering off a nucleus. Cross sections for any $\textrm{C}_i$--hadron can then be calculated. As said before, we take into account in this article only $\textrm{u}$ and $\textrm{d}$ quarks. Although it is known that a significant amount of $\textrm{s}$ quarks should in principle be present in the gluino R--hadrons (ca 15$\%$), only a small error on the cross section is made, the difference only being 6 mbarn. The amount of hadrons with two $\textrm{s}$ quarks or heavier is negligible. For a gluino R--meson and R--baryon, the total cross sections are 24 mbarn and $3/2\times 24$=36 mbarn, respectively. A gluino-gluon state can be assumed to have the same cross section as a gluino R--meson, since the geometrical cross section is approximated by the high energy hadron cross section, where gluon exchange would dominate. The gluon-gluon coupling is a factor 9/4 larger than the quark-gluon coupling, but a meson has two quarks, resulting in a cross section of a gluino-gluon state which is (9/4)/(1+1)$\approx$1 times the cross section for a gluino R--meson.
\subsection{The size of the partial cross sections}\label{sec:cross2}
\begin{figure}[t!]
\begin{center}
\epsfig{file=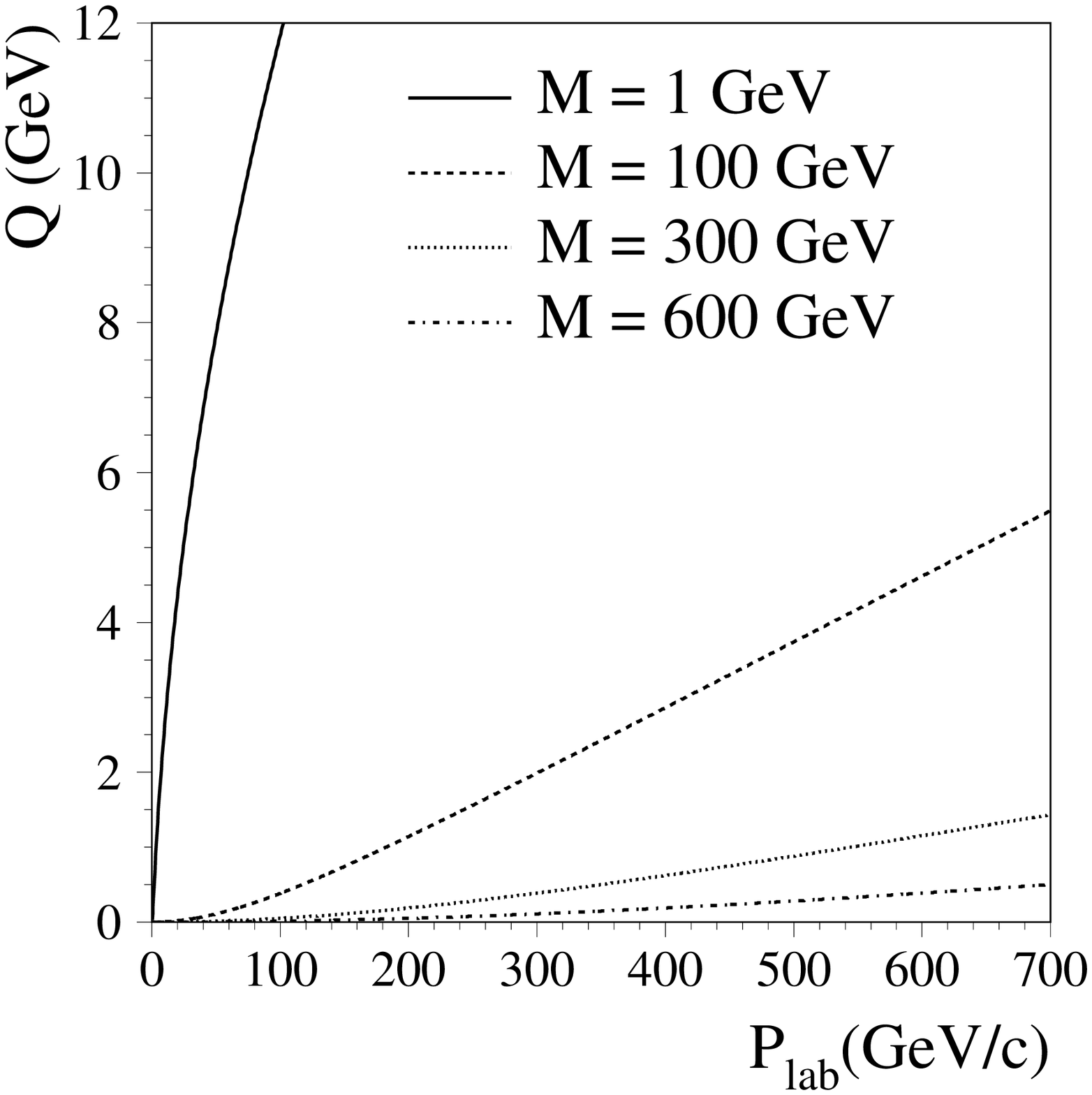,height=6.5cm,width=6.5cm}\hspace{0.8cm}\raisebox{-1.cm}{\epsfig{file=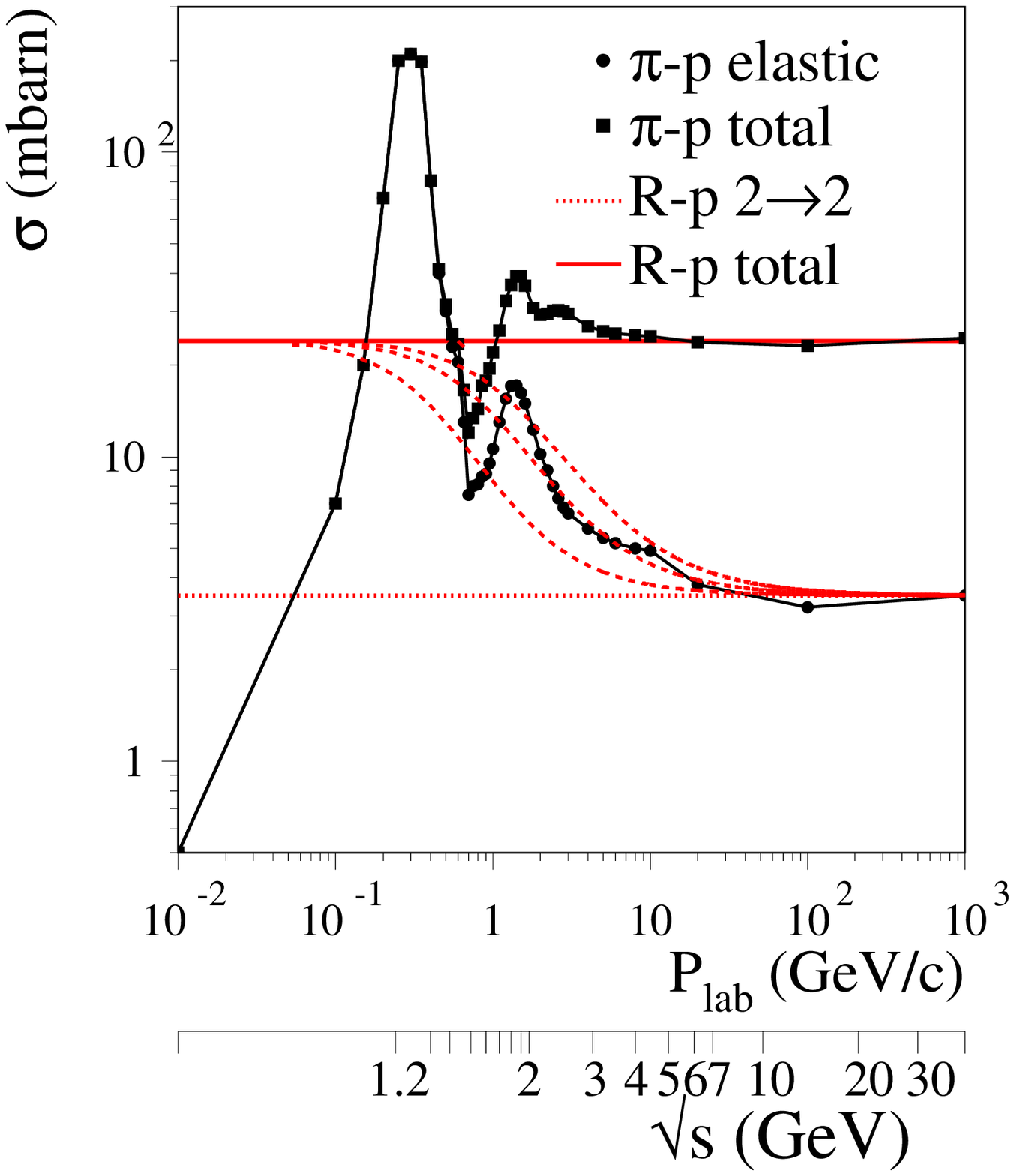,height=7.5cm,width=7.8cm}}\\
\vspace*{0.2cm}
\hspace{0.5cm}(a) \hspace{7.9cm}(b)
\end{center}
\caption{(a) The value of $Q=\sqrt{s}-m_n-m_R$, for hadrons with mass $M$ scattering with lab momentum $P_{lab}$ on a nucleon at rest. (b) The value of the cross section for gluino R--meson-proton scattering (horizontal lines) in comparison with pion-proton scattering, as well as the determination of $Q_0$ and its uncertainty (curved dashed lines connecting the horizontal lines). \label{fig:qvalue}}
\end{figure}

Besides the total cross section, which determines the interaction length, the partial cross sections for $2\rightarrow 2$ and  $2\rightarrow 3$ processes must be known in order to estimate the energy loss per nuclear interaction. An estimate of the number of final--state particles can be made by studying the energy $Q$ available for the production of kinetic energy and potentially extra particles. In the case of a $2\rightarrow N$ scattering, where an incoming particle with mass $m_1$ scatters on a particle at rest with mass $m_2$, we define the $Q$-value by
\begin{equation}\label{eq:qvalue}
Q=\sqrt{s}-m_3-m_4-...-m_{N+2}
\end{equation}where $m_i$ is the mass of final--state particle $i$. The value of $\sqrt{s}$ is $m_1^2+m_2^2+2E_1^{lab}m_2$, where $E_1^{lab}$ is the energy of the incoming particle. If the $Q$-value exceeds the mass of a pion, then an extra pion may be produced. To determine the relevant order of magnitude of $Q$ in the scattering of heavy objects, consider an elastic scattering of a heavy particle with mass $m_1$, moving with a Lorentz factor $\gamma$, off a light particle at rest with mass $m_2$. It is easily seen that $\sqrt{s}\approx m_1+\gamma m_2$ and $Q\approx (\gamma-1)m_2$, the latter being obviously small if the incoming heavy particle has a small value of $\gamma$. Fig.~\ref{fig:qvalue}a displays the $Q$-value of a process $R+n\rightarrow R+n$, in which a heavy gluino hadron $R$ with mass $M$ scatters off a nucleon $n$ with mass $m_n$ at rest.
 Fig.~\ref{fig:qvalue}a shows that for heavy hadrons, the $Q$-value is much smaller in comparison with light hadrons of the same momentum, implying that the number of final--state particles is also much smaller. For the range of momenta of particles produced at a collider like LHC, the $Q$ values are so small that the final--state multiplicity is rarely expected to exceed three. Even though the $Q$-values may exceed the pion mass, observation of for example pion--proton scattering shows that three final--state particles are produced only above $\sqrt{s}\approx$ 1.5 GeV, where $Q\approx$ 0.4 GeV $\gg m_{\pi}$. The relative amounts of the $2\rightarrow 2$ and  $2\rightarrow 3$ processes occurring for heavy hadrons at high energy can be derived from high energy pion--proton scattering data, where the cross section for $2\rightarrow 2$ and $2\rightarrow 3$ processes is known to be 15$\%$ and 85$\%$  of the total cross section, respectively. In Fig.~\ref{fig:qvalue}b, such cross sections are displayed for gluino R--meson nucleon scattering, together with the cross sections for pion--proton scattering. At low momentum, the amount of  $2\rightarrow 3$ processes is then obviously overestimated, and sometimes $2\rightarrow 3$ processes are even kinematically impossible. To solve this, a phase space function describing the relative amounts of  $2\rightarrow 2$ and  $2\rightarrow 3$ processes is introduced.
\subsection{Phase space considerations}\label{sec:ps}
In order to predict accurately the energy loss of heavy hadrons in matter, the relative amount of $2\rightarrow 2$ and  $2\rightarrow 3$ processes must be estimated. A constant matrix element is assumed for both processes, so that the relative yield of the three-body final states is initially determined solely by a phase space factor. A function describing the available phase space is expected to be a function of $Q$, defined in Eq.~\ref{eq:qvalue}, and a reasonable ansatz is given by 
\begin{equation}
F(Q)=\frac{\frac{\textrm{d}\Phi_3 (Q)}{\textrm{d}\Phi_3 (Q_0)}}{\frac{\textrm{d}\Phi_2 (Q)}{\textrm{d}\Phi_2(Q_0)}+\frac{\textrm{d}\Phi_3(Q)}{\textrm{d}\Phi_3(Q_0)}},
\end{equation}
where $\textrm{d}\Phi_n$ is the available $n$-body phase space. The fact that the two- and three-body phase spaces do not have the same dimension (compensated for by the here neglected matrix elements) forces us to normalize phase space to that available at a certain value $Q_0$. For $Q< m_{\pi}$, the phase space for three-body scattering is vanishing, while for $Q\rightarrow \infty$, it is entirely open. Simplified expressions for $\textrm{d}\Phi_2$ and $\textrm{d}\Phi_3$ can be derived under the assumption that the mass of the heavy hadron is much larger than the nucleon mass.
\begin{figure}[t!]
\begin{center}
\hspace{0.cm}\epsfig{file=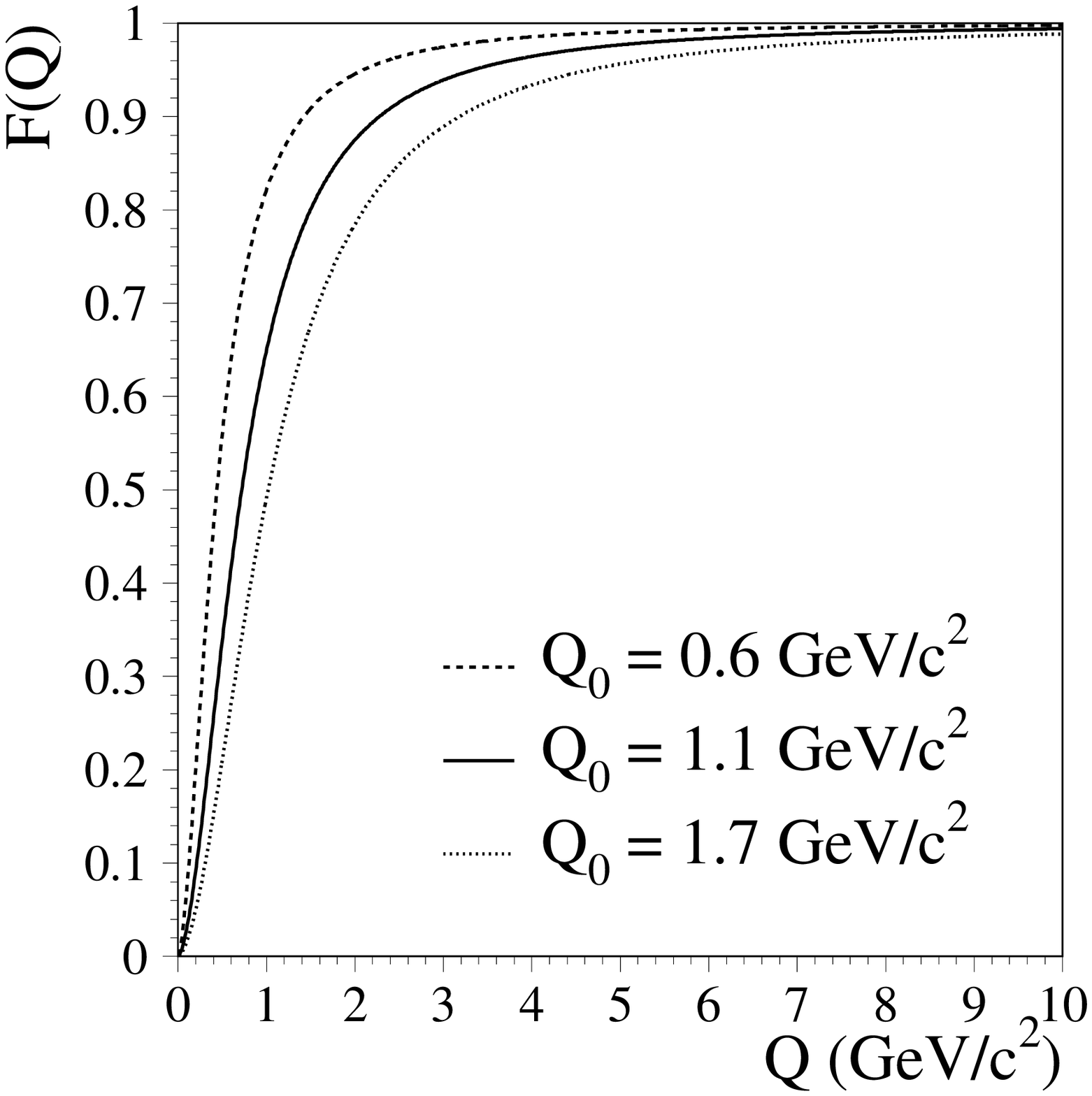,height=6.5cm,width=6.5cm}\hspace{0.8cm}\raisebox{-0.4cm}{\epsfig{file=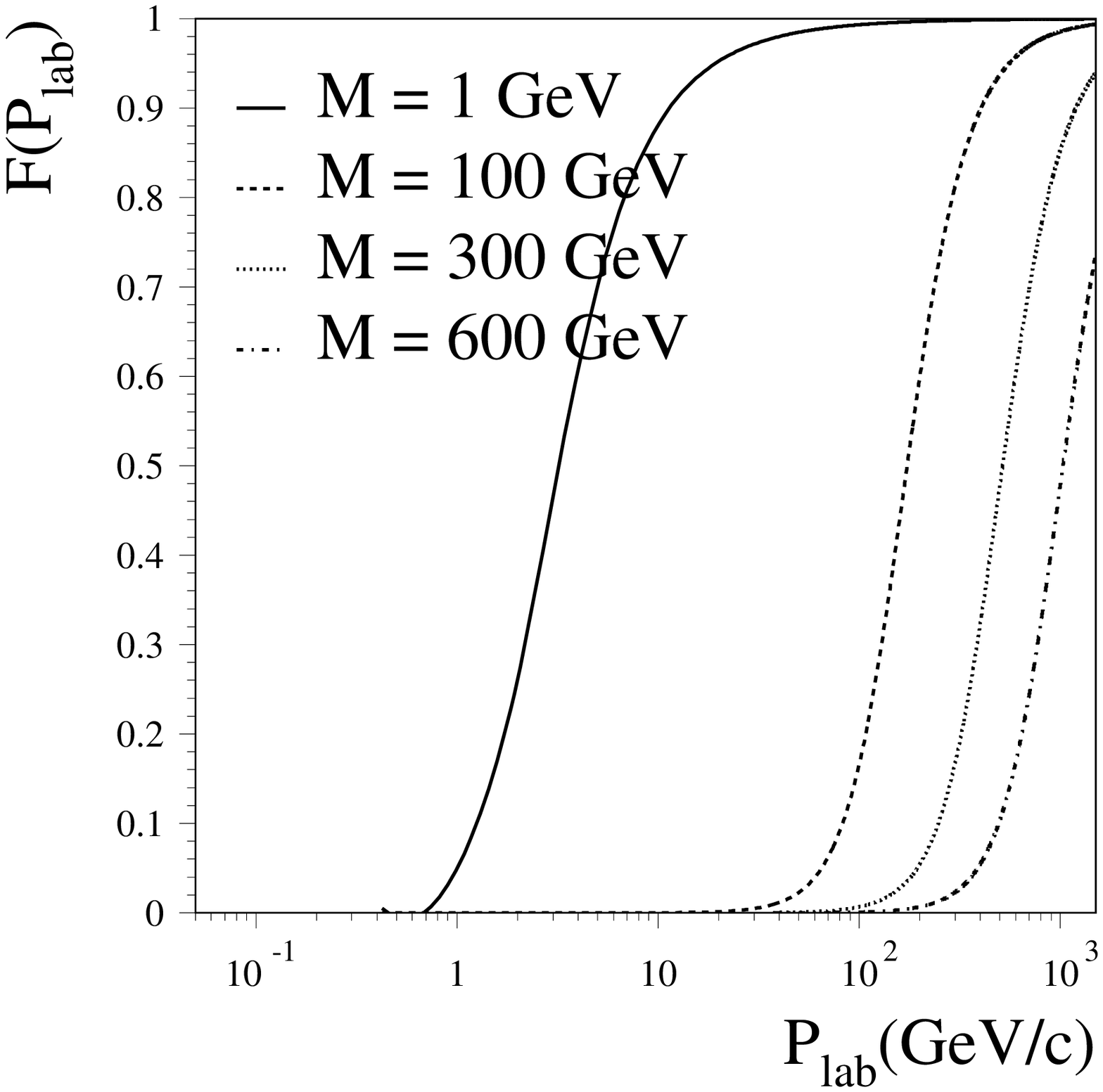,height=6.9cm,width=6.8cm}}
\end{center}
\hspace{4.5cm}(a) \hspace{6.8cm}(b)
\caption{(a) The phase space function $F(Q)$ drawn for a few values of $Q_0$  and (b) as a function of the lab momentum $\textrm{P}_{\textrm{lab}}$ of a heavy hadron with mass $M$ scattering off a nucleon at rest, and with $Q_0=1.1$ GeV.  \label{fig:phasespace}}
\end{figure} 
If one of the final--state particles in the three-body processes is a pion, a general expression describing the relative amount of three-body phase space as function of $Q$, if $Q>m_{\pi}$, may be derived:
\begin{equation}\label{eq:eps}
F(Q)=\frac{\sqrt{(1+\frac{Q}{2m_{\pi}})}(\frac{Q}{Q_o})^{3/2}}{1+\sqrt{(1+\frac{Q}{2m_{\pi}})}(\frac{Q}{Q_o})^{3/2}}
\end{equation}
From the optical theory for light hadrons, it is known that elastic scattering accounts for roughly 15$\%$ of the total cross section at high momenta, as is already mentioned in the previous section. Taking this into account, a function $f(Q)$ representing the relative probability for  $2\rightarrow 3$ scattering is
\begin{equation}\label{eq:eps2}
f(Q)=0.85F(Q)
\end{equation}
An appropriate value for $Q_0$ can be determined experimentally, e.g. by finding the value where $F(Q)$ is exactly 1/2, thus where $f(Q)=0.425$. Fig.~\ref{fig:qvalue}b illustrates the way in which we determine empirically the value for $Q_0$ and its uncertainty. Three smooth dashed curves are drawn in between the horizontal lines corresponding to the 2$\rightarrow$2 and the total cross sections. The central curve is the best estimate of the division between the elastic and inelastic cross section behaviour neglecting resonances, while the other two curves represent the uncertainty. The point in which $F(Q)$ = 1/2 corresponds to a purely elastic cross section of $(0.15+0.425)\times 24$ mb = 14 mb, which occurs at $\sqrt{s}\approx$ 1.7 GeV in the central curve, where $Q=1.7-m_p-m_{\pi} = 0.6$ GeV/$c^2$. The lower and upper limit for this point are $Q=1.4-m_p-m_{\pi}=0.4 \textrm{ GeV}/c^2$ and $Q=2.1-m_p-m_{\pi}=1.1 \textrm{ GeV}/c^2$. Substitution of $Q$ in Eq.~\ref{eq:eps} gives $Q_0$ = 1.1, with upper and lower limits of 0.6 and 1.7 GeV/$c^2$, respectively. This value roughly fits with the values obtained for $p-p$ and $p-K^+$-scattering processes. The resulting phase space function $F$ is drawn in Fig.~\ref{fig:phasespace}a, as function of $Q$, as well as the upper and lower limits. In Fig.~\ref{fig:phasespace}b, the phase space function with $Q_0$ = 1.1 GeV/$c^2$ is drawn as function of the lab momentum of the incoming hadron for different mass values.

\subsection{Nuclear scattering processes}\label{sec:proc}
As discussed in Section~\ref{sec:ref}, the presence of the gluino changes the colour-state of the hadron cloud, but not the character of the interactions of this cloud. Since scattering processes are low--energetic, all scattering processes for mesons and baryons can simply be derived by exchanging quarks. Below, interactions of hadronized colour octet and colour triplet states will be discussed.
\subsubsection{Interactions of $\textrm{C}_8\textrm{q}\bar{\textrm{q}}$}
Interactions of  $\textrm{C}_8\textrm{q}\bar{\textrm{q}}$ states include the following processes. The $2\rightarrow 2$ processes are purely elastic scattering (e.g.  $\textrm{C}_8\textrm{d}\bar{\textrm{d}}+\textrm{u}\textrm{u}\textrm{d} \rightarrow \textrm{C}_8\textrm{d}\bar{\textrm{d}}+ \textrm{u}\textrm{u}\textrm{d}$), charge exchange (e.g. $\textrm{C}_8\textrm{d}\bar{\textrm{d}}+\textrm{u}\textrm{u}\textrm{d} \rightarrow \textrm{C}_8\textrm{u}\bar{\textrm{d}} + \textrm{u}\textrm{d}\textrm{d}$) and baryon exchange (e.g. $\textrm{C}_8\textrm{d}\bar{\textrm{d}}+\textrm{u}\textrm{u}\textrm{d} \rightarrow \textrm{C}_8\textrm{u}\textrm{d}\textrm{d} + \textrm{u}\bar{\textrm{d}}$) while the $2\rightarrow 3$ processes include normal inelastic scattering (e.g. $\textrm{C}_8\textrm{d}\bar{\textrm{d}}+  \textrm{u}\textrm{u}\textrm{d} \rightarrow \textrm{C}_8\textrm{u}\bar{\textrm{d}}+ \textrm{u}\textrm{d}\textrm{d} + \textrm{d}\bar{\textrm{d}}$) and inelastic scattering with baryon exchange (e.g. $\textrm{C}_8\textrm{d}\bar{\textrm{d}} + \textrm{u}\textrm{u}\textrm{d} \rightarrow \textrm{C}_8\textrm{u}\textrm{u}\textrm{d} + \textrm{u}\bar{\textrm{d}} + \textrm{d}\bar{\textrm{u}}$). 
 
\noindent It should be noted that the processes with baryon exchange are kinematically favoured due to the fact that a final-state pion is so light. In such processes, typically an extra $2m_{d}-m_{\pi}\approx$ 500 MeV of kinetic energy would be liberated. However, these processes could be dynamically suppressed because two quarks must be exchanged. 
\subsubsection{Interactions of $\textrm{C}_8\textrm{q}\textrm{q}\textrm{q}$ states.}
Interaction processes include $2\rightarrow 2$ processes like purely elastic scattering (e.g.  $\textrm{C}_8\textrm{u}\textrm{u}\textrm{d}+\textrm{u}\textrm{u}\textrm{d} \rightarrow \textrm{C}_8\textrm{u}\textrm{u}\textrm{d}+ \textrm{u}\textrm{u}\textrm{d}$) and charge exchange (e.g. $\textrm{C}_8\textrm{u}\textrm{u}\textrm{d}+\textrm{u}\textrm{d}\textrm{d} \rightarrow \textrm{C}_8\textrm{u}\textrm{d}\textrm{d} + \textrm{u}\textrm{u}\textrm{d}$) and $2\rightarrow 3$ processes like $\textrm{C}_8\textrm{u}\textrm{u}\textrm{d}+  \textrm{u}\textrm{d}\textrm{d} \rightarrow \textrm{C}_8\textrm{u}\textrm{d}\textrm{d}+ \textrm{u}\textrm{u}\textrm{d} + \textrm{d}\bar{\textrm{d}}$. No baryon exchange is possible, since the probability that the $\textrm{C}_8\textrm{q}\textrm{q}\textrm{q}$ state interacts with a pion in the nucleus is negligible, and besides this process would kinematically be strongly disfavoured. This has the important implication that mesons get converted into baryons during repeated interactions, but not vice versa.

\subsubsection{Interactions of $\textrm{C}_8\bar{\textrm{q}}\bar{\textrm{q}}\bar{\textrm{q}}$ states.} 
Only a very small amount of $\textrm{C}_8\bar{\textrm{q}}\bar{\textrm{q}}\bar{\textrm{q}}$ states will arise in the hadronization process. Interactions differ from those of $\textrm{C}_8\textrm{q}\textrm{q}\textrm{q}$ states, since scattering takes place on protons or neutrons, containing quarks rather than antiquarks. A $\textrm{C}_8\bar{\textrm{q}}\bar{\textrm{q}}\bar{\textrm{q}}$ state would thus interact dominantly by baryon annihilation, for example $\textrm{C}_8\bar{\textrm{u}}\bar{\textrm{u}}\bar{\textrm{d}}+\textrm{u}\textrm{u}\textrm{d}\rightarrow \textrm{C}_8\textrm{u}\bar{\textrm{d}}+\textrm{u}\bar{\textrm{d}}$. This process would kinematically be favourable.    
\subsubsection{Interactions of $\textrm{C}_8\textrm{g}$ states}
A gluon is able to convert into a $\textrm{u}\bar{\textrm{u}}$ or $\textrm{d}\bar{\textrm{d}}$ state, and as such, a $\textrm{C}_8\textrm{g}$ state  probably interacts like (and mixes with) $\textrm{C}_8\textrm{u}\bar{\textrm{u}}$ or $\textrm{C}_8\textrm{d}\bar{\textrm{d}}$ states. The mass of the active system, the gluon, is usually taken to be 0.7 GeV~\cite{Pennington:1998ys}, approximately the same as the constituent mass of two first generation quarks. Thus, the possible interaction processes are expected to be similar to those for  $\textrm{C}_8\textrm{u}\bar{\textrm{u}}$ or $\textrm{C}_8\textrm{d}\bar{\textrm{d}}$ states. Besides the fact that the interaction processes are similar, the cross section is expected to be roughly the same, as has already been explained in Section~\ref{sec:cross}, allowing us to treat a $\textrm{C}_8\textrm{g}$ state like a neutral $\textrm{C}_8\textrm{q}\bar{\textrm{q}}$ state. As for $\textrm{C}_8\textrm{q}\bar{\textrm{q}}$ states, the $\textrm{C}_8\textrm{g}$ states will eventually convert into baryons. 

\subsubsection{Interactions of $\textrm{C}_3\bar{\textrm{q}}$ and $\textrm{C}_{\bar{3}}\textrm{q}$ states}
Interactions of a $\textrm{C}_3\bar{\textrm{q}}$ state include processes with quark--antiquark annihilation. Thus, there are $2\rightarrow 2$ processes such as elastic scattering (e.g.  $\textrm{C}_3\bar{\textrm{u}}+\textrm{u}\textrm{u}\textrm{d} \rightarrow \textrm{C}_3\bar{\textrm{u}}+ \textrm{u}\textrm{u}\textrm{d}$), charge exchange (e.g $\textrm{C}_3\bar{\textrm{d}}+\textrm{u}\textrm{d}\textrm{d} \rightarrow \textrm{C}_3\bar{\textrm{u}} + \textrm{u}\textrm{u}\textrm{d}$),  or baryon exchange ($\textrm{C}_3\bar{\textrm{u}}+\textrm{u}\textrm{d}\textrm{d}\rightarrow \textrm{C}_3\textrm{u}\textrm{d}+\bar{\textrm{u}}\textrm{d})$. For $2\rightarrow 3$ processes, these include normal inelastic processes, like $\textrm{C}_3\bar{\textrm{u}} +  \textrm{u}\textrm{d}\textrm{d} \rightarrow \textrm{C}_3\bar{\textrm{d}} + \textrm{u}\textrm{d}\textrm{d} + \bar{\textrm{u}}\textrm{d}$, as well as processes with baryon exchange, like $\textrm{C}_3\bar{\textrm{u}}+\textrm{u}\textrm{d}\textrm{d}\rightarrow \textrm{C}_3\textrm{u}\textrm{d}+ \bar{\textrm{u}}\textrm{d}+ \textrm{u}\bar{\textrm{u}}$. As above, baryon exchange is kinematically favoured by the possibility of having a light pion in the final state.

\noindent Interactions of a $\textrm{C}_{\bar{3}}\textrm{q}$ state include similar processes but without quark--antiquark annihilation and without baryon exchange. Possible processes in that case are $2\rightarrow 2$ processes like elastic scattering (e.g.  $\textrm{C}_{\bar{3}}\textrm{u}+\textrm{u}\textrm{u}\textrm{d} \rightarrow \textrm{C}_{\bar{3}}\textrm{u}+ \textrm{u}\textrm{u}\textrm{d}$), charge exchange (e.g. $\textrm{C}_{\bar{3}}\textrm{d}+\textrm{u}\textrm{d}\textrm{d} \rightarrow \textrm{C}_{\bar{3}}\textrm{u} + \textrm{u}\textrm{u}\textrm{d}$), and $2\rightarrow 3$ processes like $\textrm{C}_{\bar{3}}\textrm{u} +  \textrm{u}\textrm{d}\textrm{d} \rightarrow \textrm{C}_{\bar{3}}\textrm{d} + \textrm{u}\textrm{u}\textrm{d} + \bar{\textrm{d}}\textrm{d}$. 
\subsubsection{Interactions of $\textrm{C}_3\textrm{q}\textrm{q}$ and $\textrm{C}_{\bar{3}}\bar{\textrm{q}}\bar{\textrm{q}}$ states }
 Interactions of $\textrm{C}_3\textrm{q}\textrm{q}$ states are processes without quark annihilation and include $2\rightarrow 2$ processes with purely elastic scattering (e.g. $\textrm{C}_3\textrm{u}\textrm{u}+\textrm{u}\textrm{u}\textrm{d} \rightarrow \textrm{C}_3\textrm{u}\textrm{u}+ \textrm{u}\textrm{u}\textrm{d}$) and charge exchange (e.g. $\textrm{C}_3\textrm{u}\textrm{u}+\textrm{u}\textrm{d}\textrm{d} \rightarrow \textrm{C}_3\textrm{u}\textrm{d} + \textrm{u}\textrm{u}\textrm{d}$), and $2\rightarrow 3$ processes like e.g. $\textrm{C}_3\textrm{u}\textrm{u} +  \textrm{u}\textrm{d}\textrm{d} \rightarrow \textrm{C}_3\textrm{u}\textrm{d} + \textrm{u}\textrm{d}\textrm{d} + \textrm{u}\bar{\textrm{d}}$. 

\noindent On the other hand, $\textrm{C}_{\bar{3}}\bar{\textrm{q}}\bar{\textrm{q}}$ states may interact by quark annihilation and thereby baryon annihilation. Processes include $2 \rightarrow 2$  processes like purely elastic scattering (e.g. $\textrm{C}_{\bar{3}}\bar{\textrm{u}}\bar{\textrm{u}}+\textrm{u}\textrm{u}\textrm{d} \rightarrow \textrm{C}_{\bar{3}}\bar{\textrm{u}}\bar{\textrm{u}}+ \textrm{u}\textrm{u}\textrm{d}$), charge exchange (e.g. $\textrm{C}_{\bar{3}}\bar{\textrm{u}}\bar{\textrm{u}}+\textrm{u}\textrm{u}\textrm{d} \rightarrow \textrm{C}_{\bar{3}}\bar{\textrm{u}}\bar{\textrm{d}} + \textrm{u}\textrm{d}\textrm{d}$), and baryon annihilation (e.g. $\textrm{C}_{\bar{3}}\bar{\textrm{u}}\bar{\textrm{u}}+\textrm{u}\textrm{u}\textrm{d} \rightarrow \textrm{C}_{\bar{3}}\textrm{u} + \bar{\textrm{u}}\textrm{d}$) and $2\rightarrow 3$ processes like e.g. $\textrm{C}_{\bar{3}}\bar{\textrm{u}}\bar{\textrm{u}} +  \textrm{u}\textrm{d}\textrm{d} \rightarrow \textrm{C}_{\bar{3}}\bar{\textrm{u}}\bar{\textrm{d}} + \textrm{u}\textrm{d}\textrm{d} + \bar{\textrm{u}}\textrm{d}$. Baryon annihilation would kinematically be favoured.

\subsection{Relative probabilities of scattering processes}
In general, an enormous number of scattering processes is possible, over 140 when summed over all gluino R--hadrons. To know which processes take place, the target (neutron or proton) must be known, as well as the relative coupling of all processes. The latter requires the calculation of the Clebsch-Gordan coefficients of isospin-related processes, plus an evaluation of all additional dynamical effects for all processes. This however has not been done. Since the masses of the different lowest-lying $\textrm{C}_8$ mesons are degenerate, and similarly for the baryons, as discussed in Section~\ref{hadrons}, the phenomenology is not expected to be affected by this assumption. Therefore, after the target and the class of interaction processes, i.e.~$2\rightarrow 2$ or $2\rightarrow 3$ processes, is determined, equal relative couplings for all processes inside this class are assumed. 
\subsection{The nuclear cascade}
The incoming hadron interacts with a nucleon inside a nucleus. Issues like the Fermi motion of the nucleons inside the nucleus, binding energy of the nucleus, evaporation energy of a nucleus, and instability of the nucleus after the interaction play an important role. It depends on the energy of the incident hadron whether the interaction causes a large nuclear cascade or not. To estimate this, the wavelength associated with the incident hadron, $\lambda=\frac{hc}{E}$, can be compared with the size of the nucleus. Since the interacting system of a heavy hadron is low--energetic, the associated wavelengths are of the order of the size of the nucleus, and the development of a nuclear cascade is probable.

\subsection{Nuclear energy losses}
As argued in Section~\ref{sec:cross2} the total energy loss in a collision of the heavy hadron is small. This small energy transfer is shared between the kinetic energy of the kicked-out nucleon or pion, its binding energy in the nucleus, the production of extra final--state particles, and nuclear degrees of freedom inside the remnant.

\begin{figure}[t!]
\begin{center}
\epsfig{file=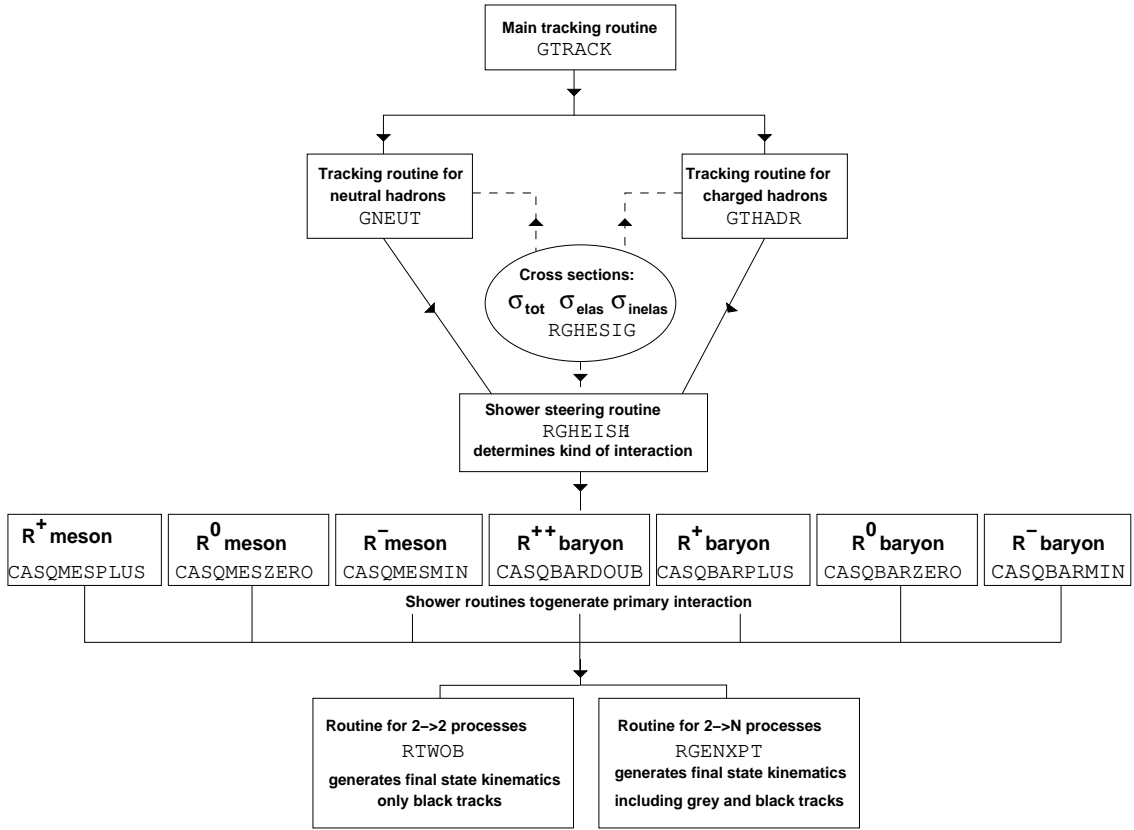,height=11.cm,width=15cm}
\end{center}
\caption{The tracking steps in GHEISHA. \label{fig:trackback}}
\end{figure}
\section{GEANT 3 simulation of R--hadronic interactions}\label{sec:geant}The simulation of processes which accompany the propagation of an R--hadron through the detector material is performed within the framework of GEANT 3. For the specific case of describing R--hadronic interactions in matter, the GHEISHA package is used~\cite{gheishamanual}. Dedicated routines to describe interactions of gluino R--hadrons are developed~\cite{aafke}. Tracking of R--hadrons through a volume is done without modification. For the generation of R--hadronic nuclear interactions, the existing pion routines are used as starting point, following the idea of Refs.~\cite{Barate:2000qf} and~\cite{rhadrons1}. In Fig.~\ref{fig:trackback}, the GHEISHA tracking steps and the new routines are displayed. Three main issues characterize the generation of hadronic interactions in GHEISHA, which are simulated as follows.
\begin{enumerate}
\item{The evaluation of the cross sections, needed to calculate the mean free path. Cross sections are calculated according to the arguments in Section~\ref{sec:cross}.}
\item{The selection of the interaction and subsequent sampling of the final--state multiplicities. The target and selection of the interaction process is done with the help of seven new shower routines for gluino $R^{++}, R^{+}, R^0$ and $R^-$ baryons and for $R^{+}, R^0$ and $R^-$ mesons. Due to the tiny amount of $\textrm{C}_8\bar{\textrm{q}}\bar{\textrm{q}}\bar{\textrm{q}}$ states no shower routine routines devoted to this states have been developed. The interaction processes are chosen according to the prescription in Section~\ref{sec:ref4}.}

\item{The generation of the final--state particles and their kinematics. The $p_T$-values for the final--state particles are tabulated from experimental data. For R--hadrons, this implies that $t$ values are selected as in the pion routines following d$\sigma$/d$t$=$e^{-bt}$, where $b$ is an empirical function of the lab momentum of the incoming hadron. A rescaling of momenta to the momenta of the active quark (or gluon) system is applied for R--hadrons. }
\end{enumerate}
The fact that the hadron scatters on a nucleus and not on a free proton is taken into consideration in the three issues mentioned above. Apart from a rescaling of the evaporation energy of the nucleus, being a function of the kinetic energy of the incoming hadron, to the kinetic energy of the active incoming quark (or gluon) system, the issues are simulated exactly as for pions. The code, as well as more detailed information about the simulation, is available in Ref.~\cite{aafke}. 
\section{Results}\label{sec:pros}
\subsection{Meson-baryon conversion}
\begin{figure}[t!]
\begin{center}
\epsfig{file=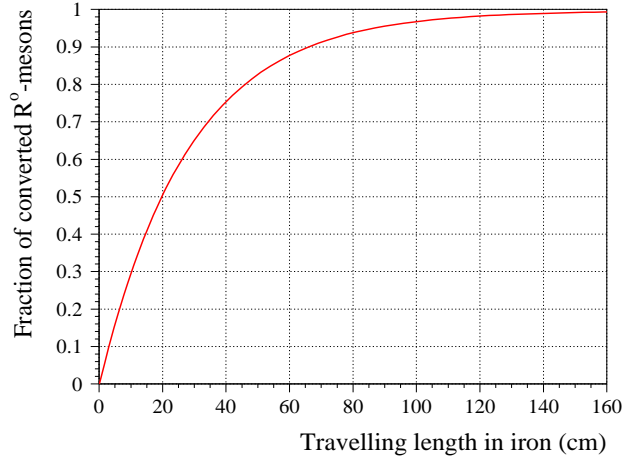,height=7cm,width=9cm}
\end{center}
\caption{The fraction of R--mesons converted into R--baryons as function of the traveling length, starting with a neutral R--meson.  \label{fig:conversion}}
\end{figure}

As explained in detail above, an R--meson is able to convert into an R--baryon by scattering off a nucleon, but not vice-versa. The amount of converted mesons as a function of the traveling length in a piece of iron is shown in Fig.~\ref{fig:conversion}. Note that an R--baryon has a larger cross section for nuclear interactions than an R--meson. The conversion therefore increases the subsequent energy losses in a calorimeter.
\begin{figure}[b!]
\begin{center}
\epsfig{file=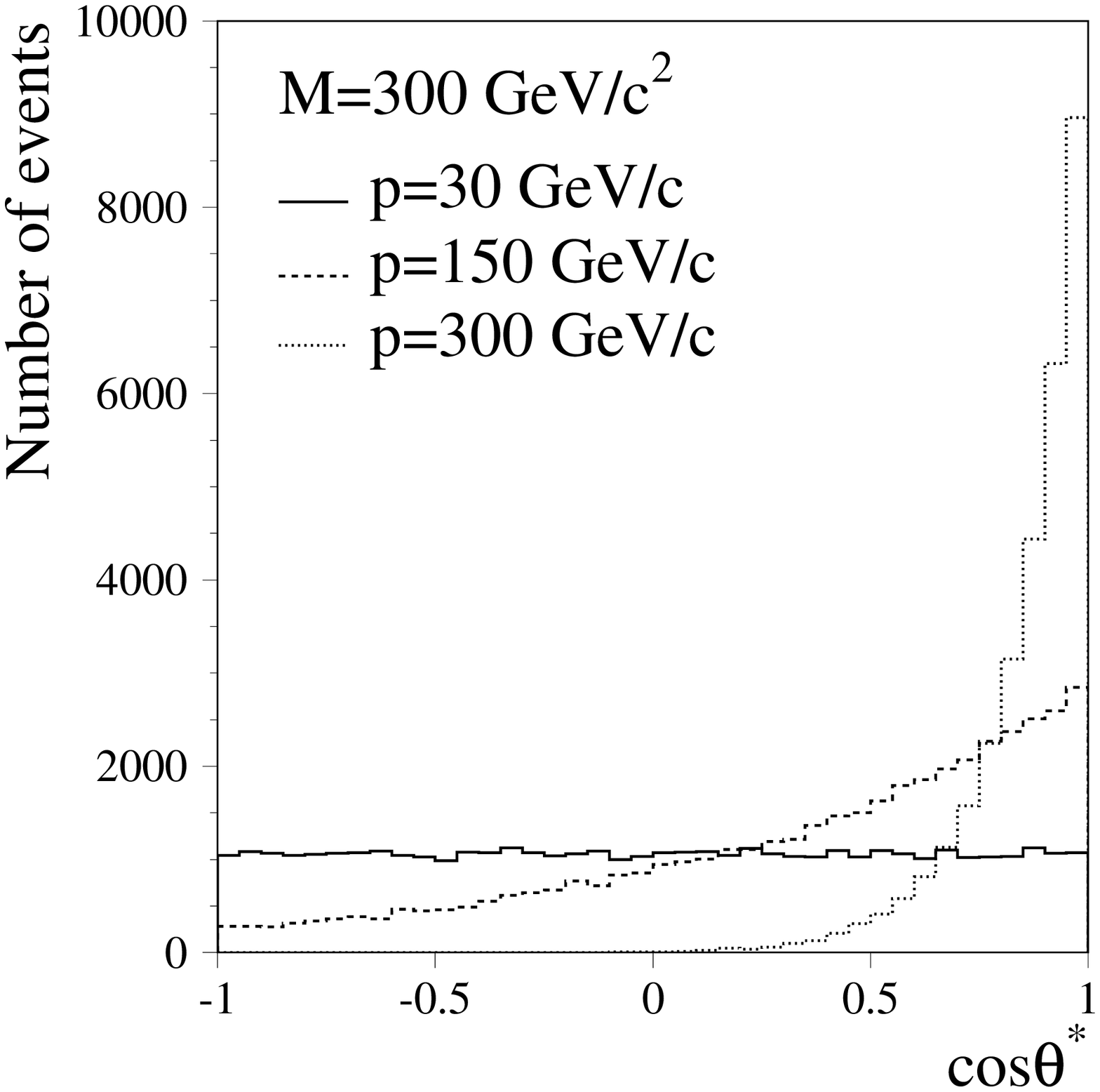,height=4.6cm,width=4.3cm}\epsfig{file=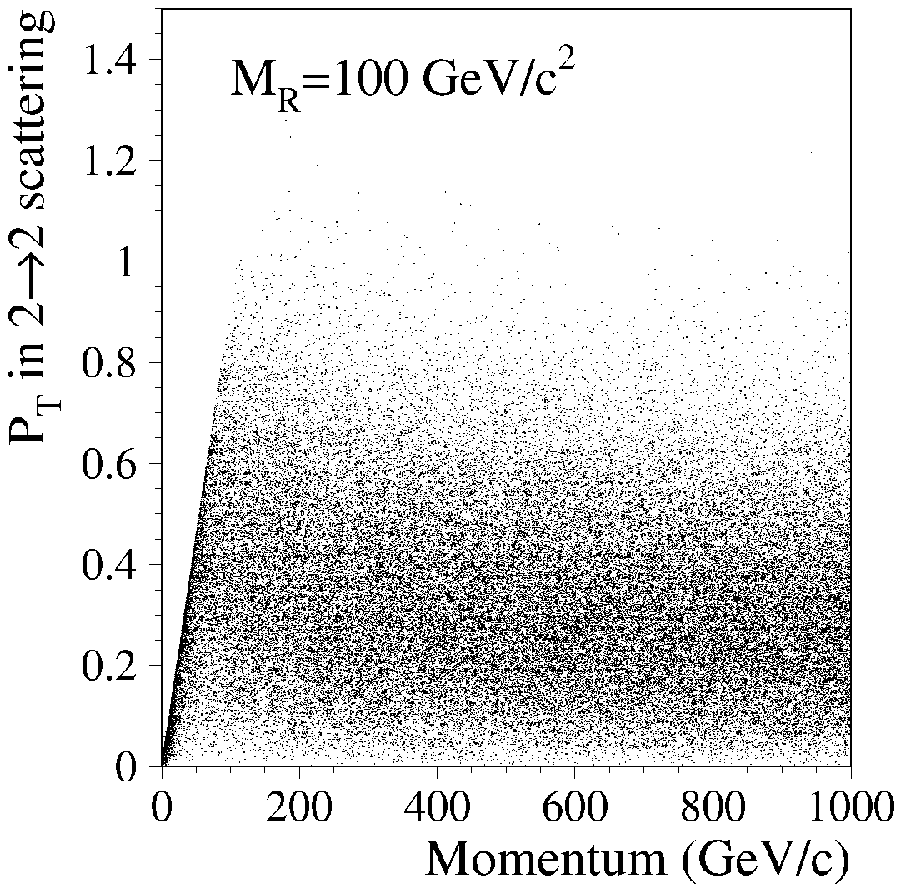,height=4.6cm,width=4.3cm}\epsfig{file=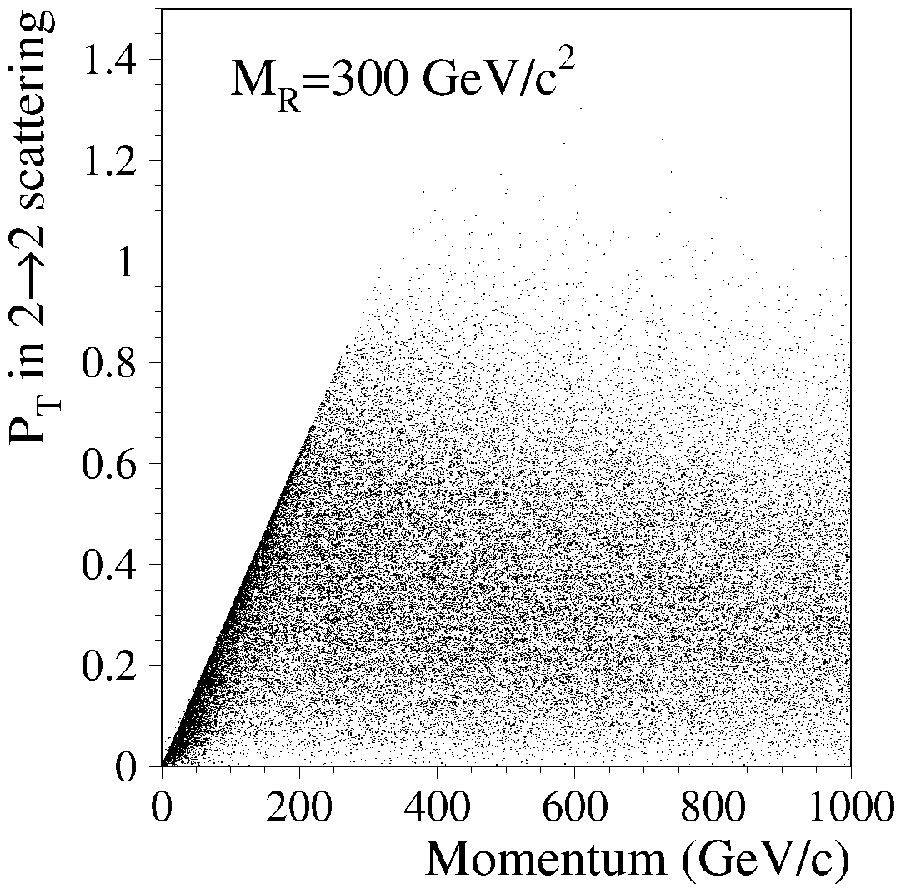,height=4.6cm,width=4.3cm}\epsfig{file=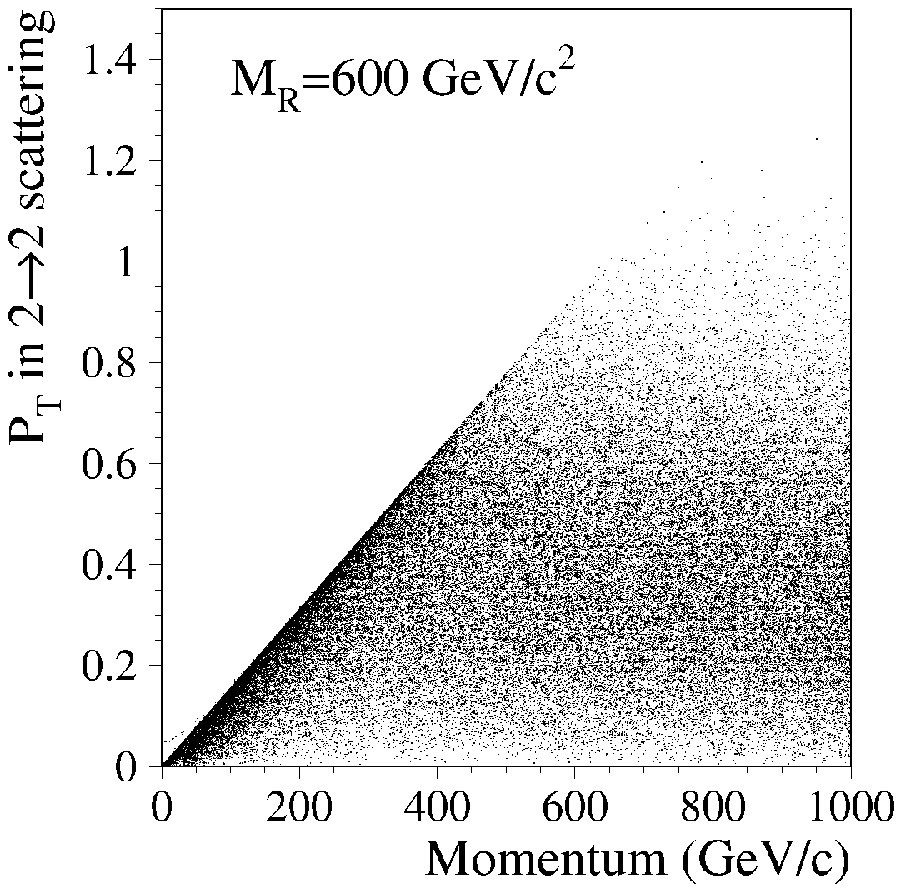,height=4.6cm,width=4.3cm}
\end{center}
\caption{At the left, distributions for $\cos\theta^*$ in 2$\rightarrow$2 scattering for different values of the R--hadron initial momentum (left). At the right, scattering plots for the generated $p_T$ value of the R--hadron in 2$\rightarrow$2 scattering for three different values of the R--hadron mass: 100, 300 and 600 GeV/$c^2$. \label{fig:tandcost}}
\end{figure}
\subsection{Kinematics}
We cross-check the values for the scattering angle in the center of mass system, $\cos\theta^*$, and for the transverse momentum $p_T$ of the scattered R--hadron. Results are shown in Fig.~\ref{fig:tandcost}. At small momenta, the $2\rightarrow 2$ scattering processes are isotropic, while at large momentum, a clear forward peak is seen, explained by the fact that the $p_T$ in nuclear scatterings is limited to approximately 1 GeV.
\begin{figure}[t!]
\begin{center}
\epsfig{file=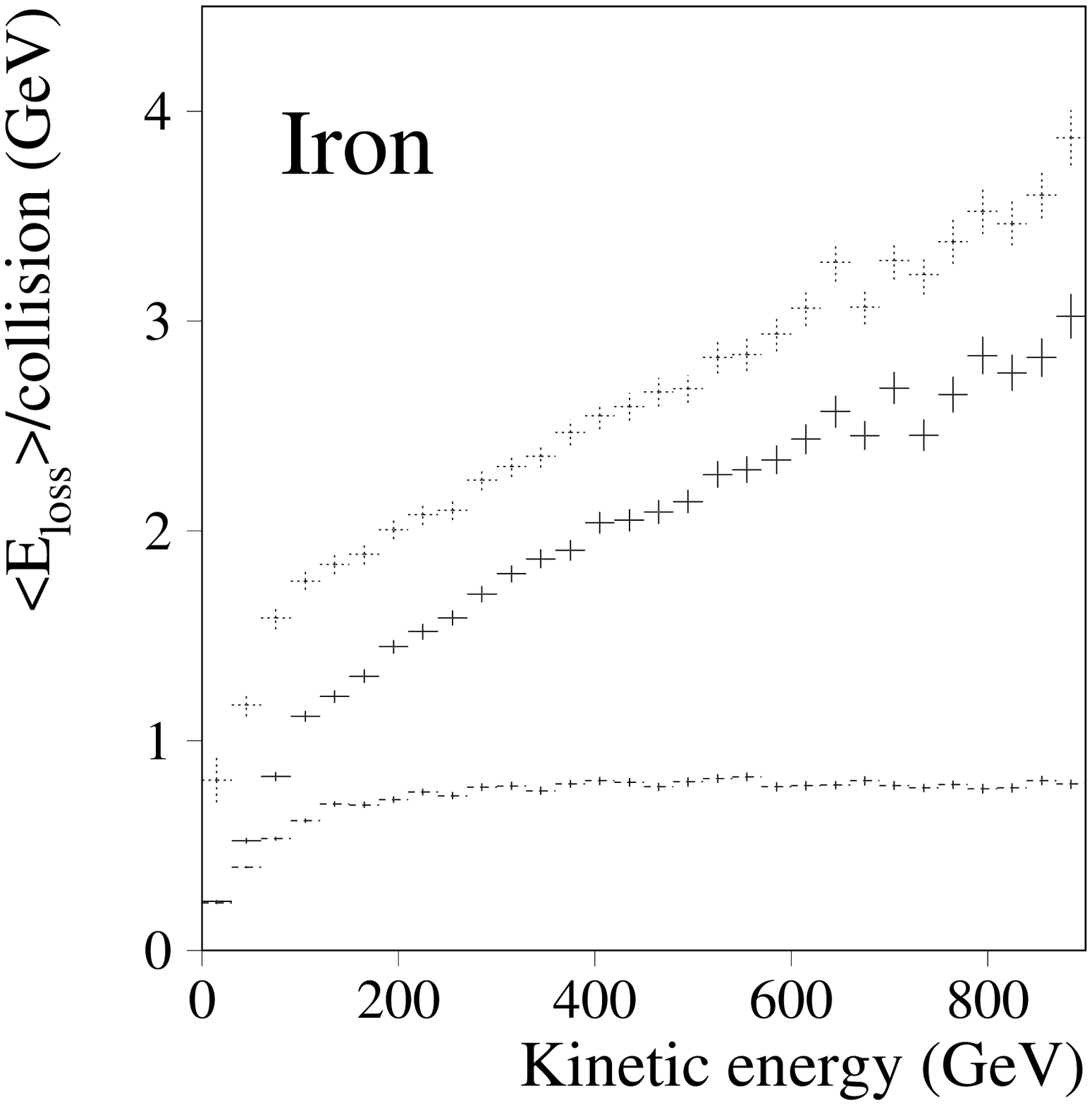,height=5.1cm,width=5.1cm}\hspace{0.4cm}\epsfig{file=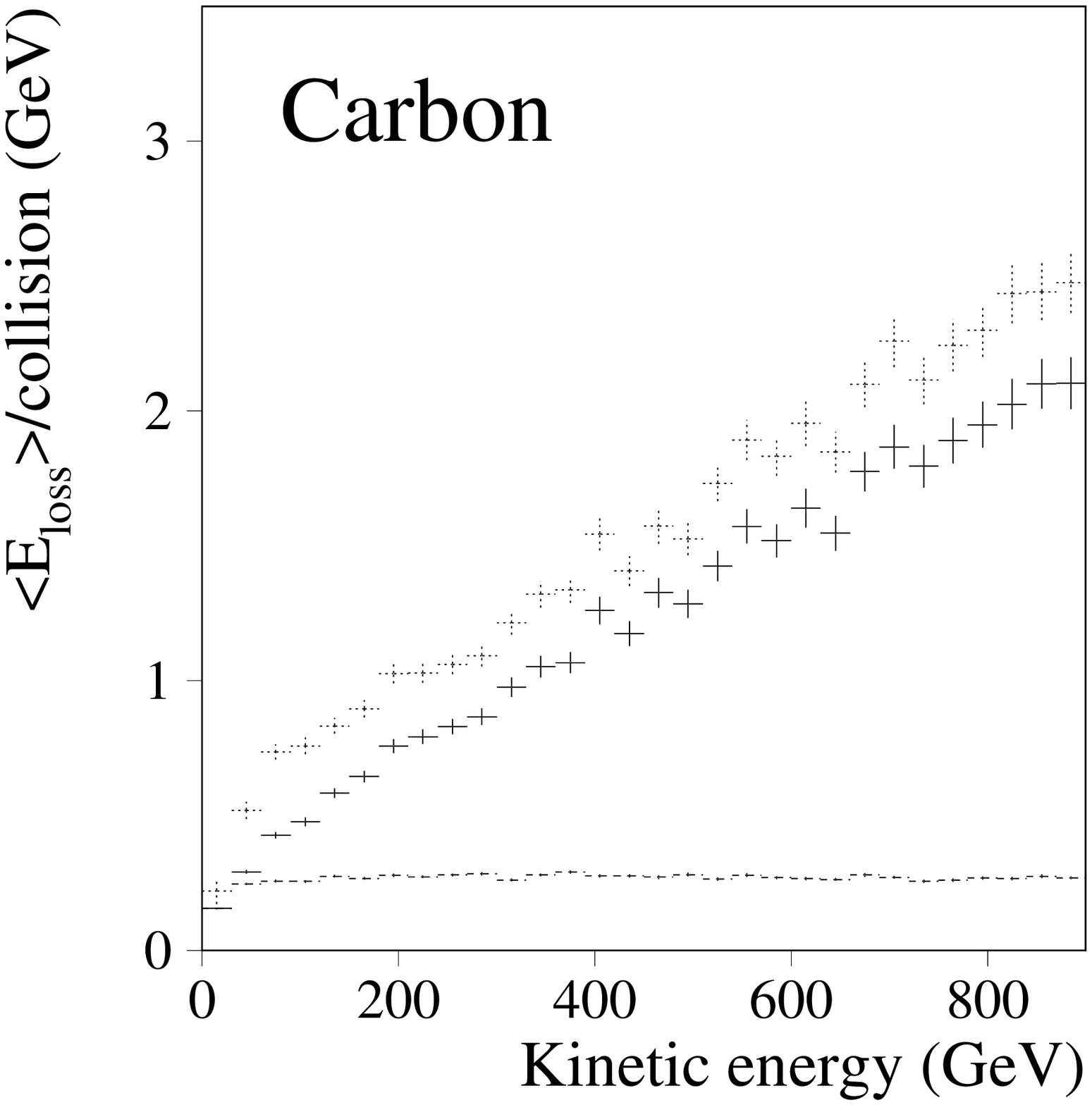,height=5.1cm,width=5.1cm}\hspace{0.4cm}\epsfig{file=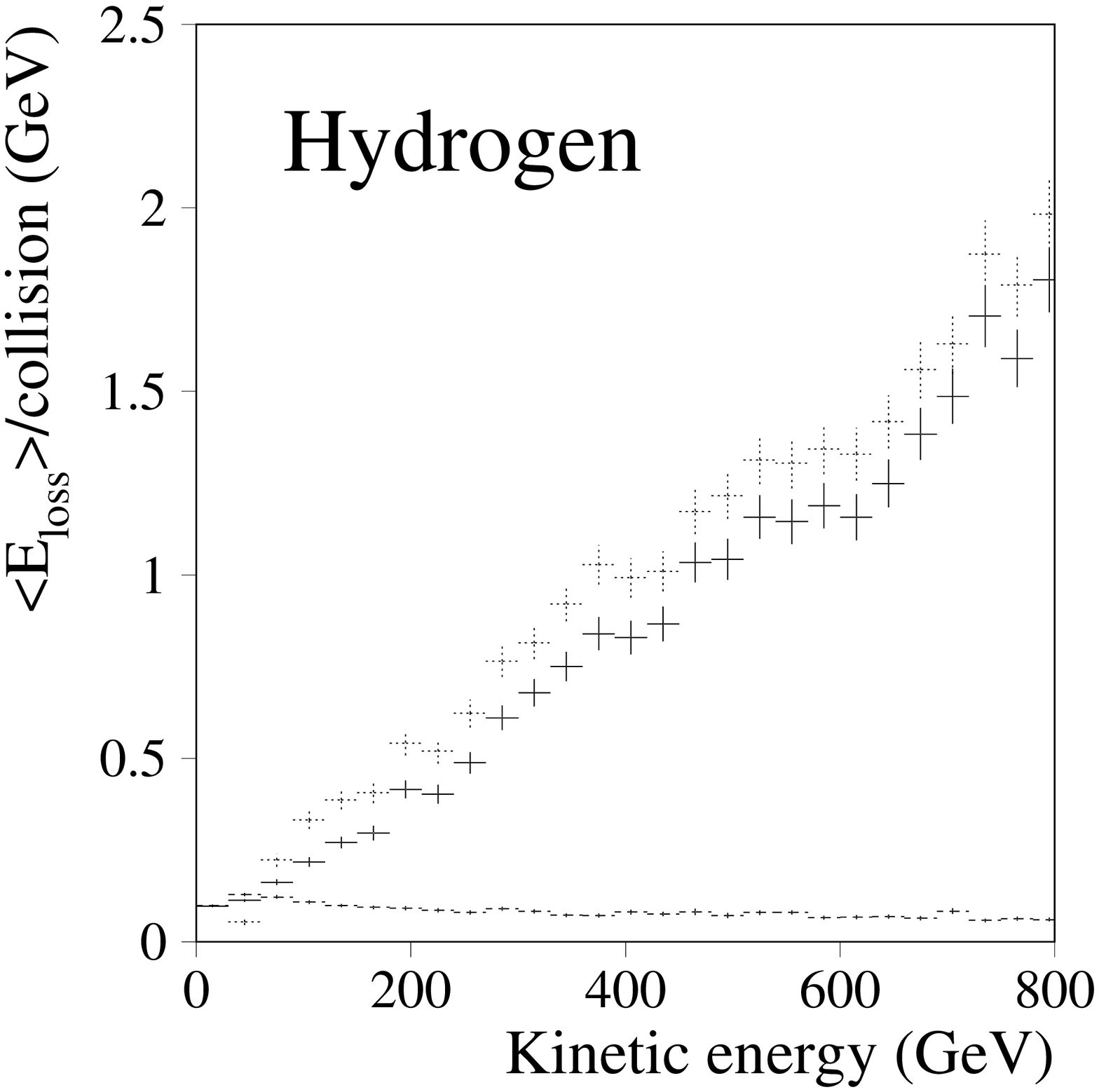,height=5.1cm,width=5.1cm}
\epsfig{file=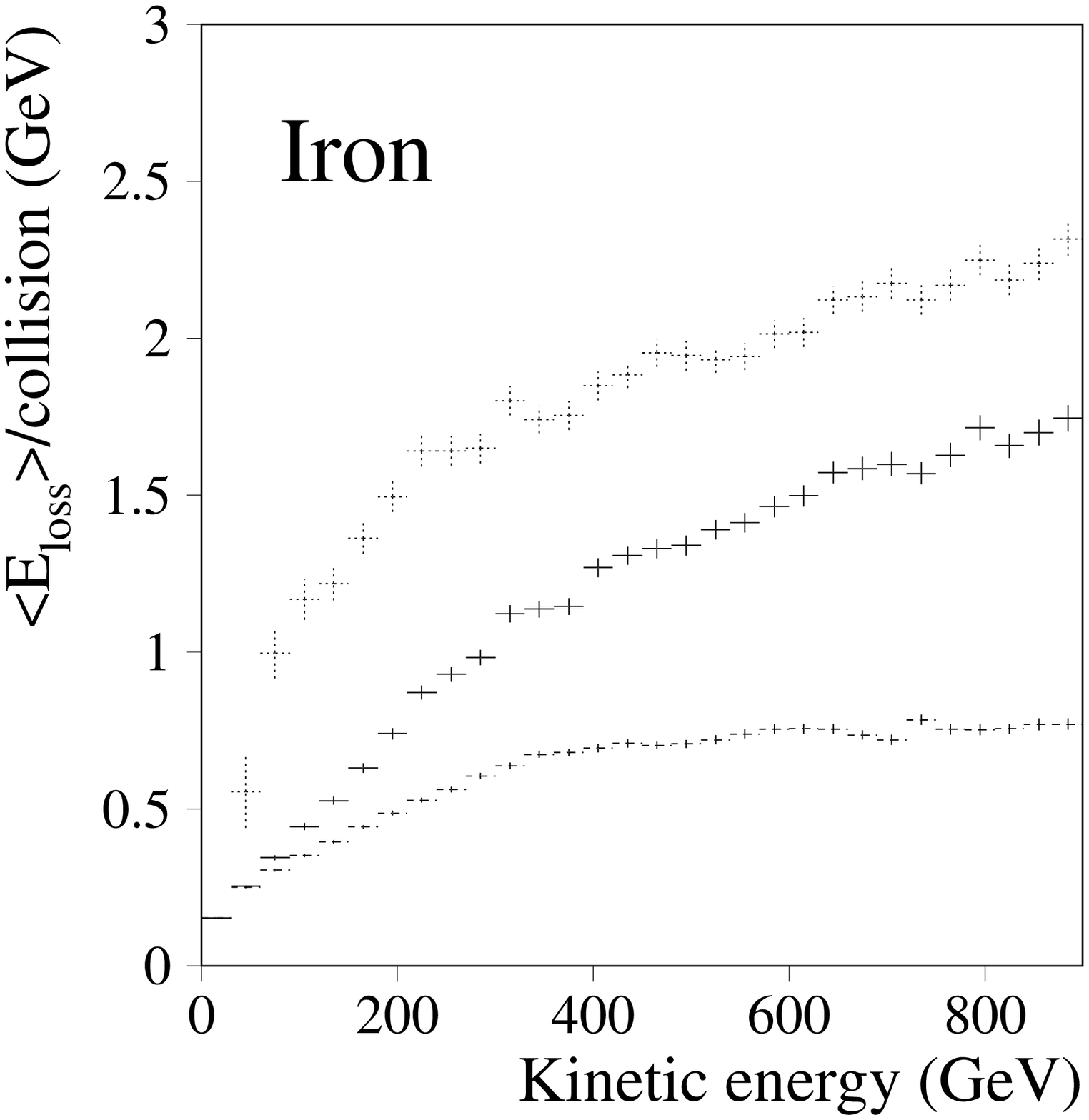,height=5.1cm,width=5.1cm}\hspace{0.4cm}\epsfig{file=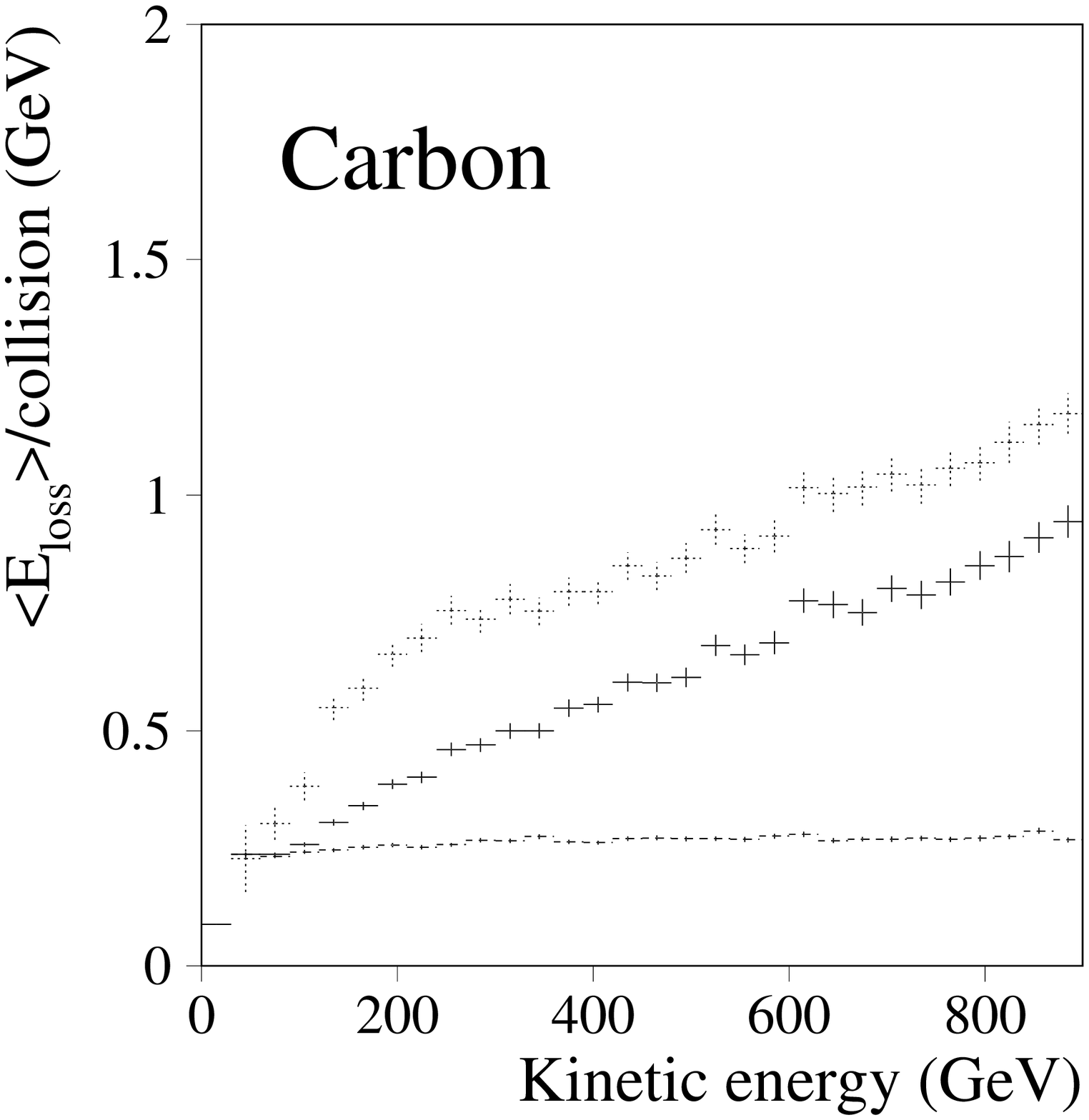,height=5.1cm,width=5.1cm}\hspace{0.4cm}\epsfig{file=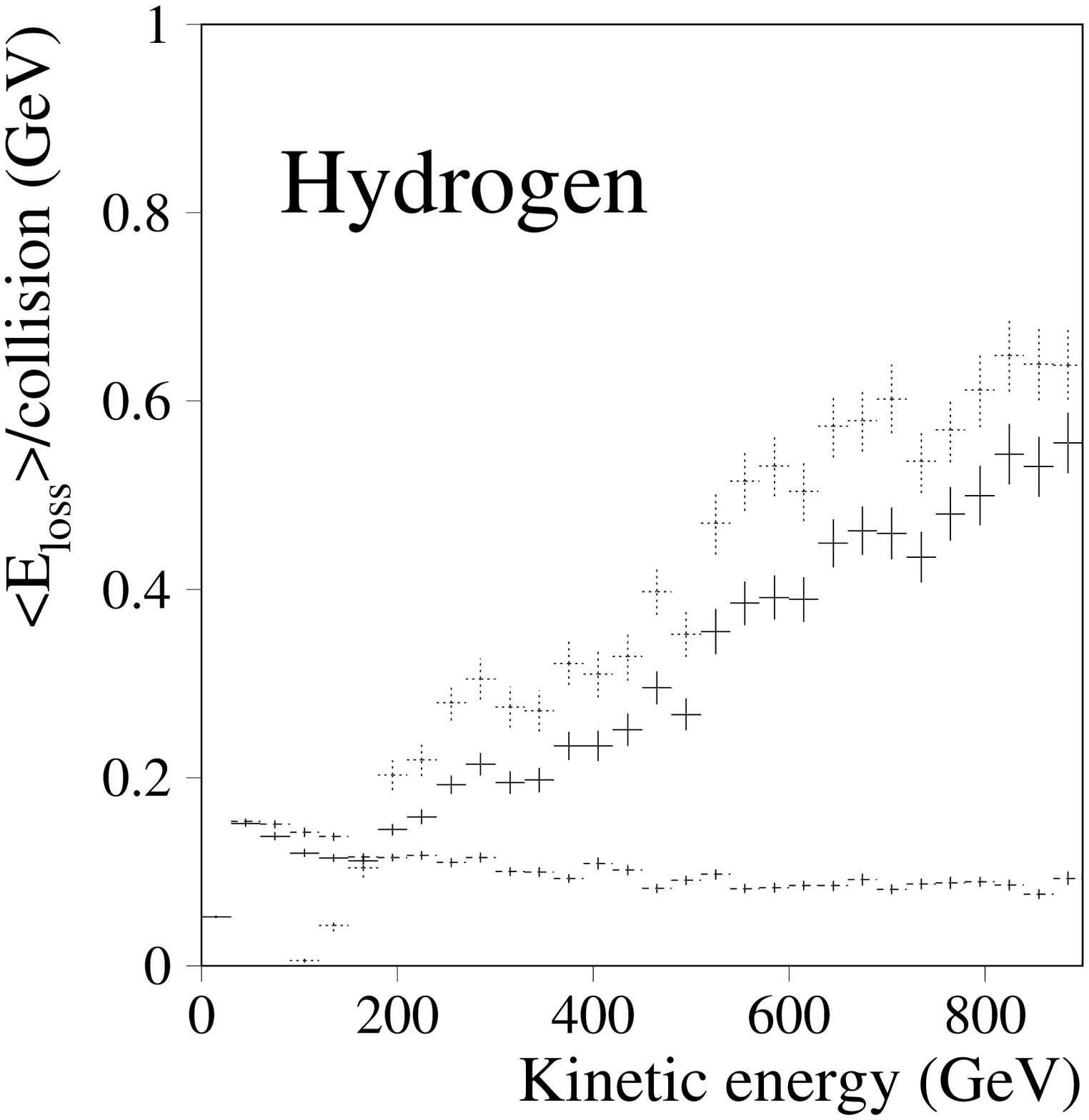,height=5.1cm,width=5.1cm}
\epsfig{file=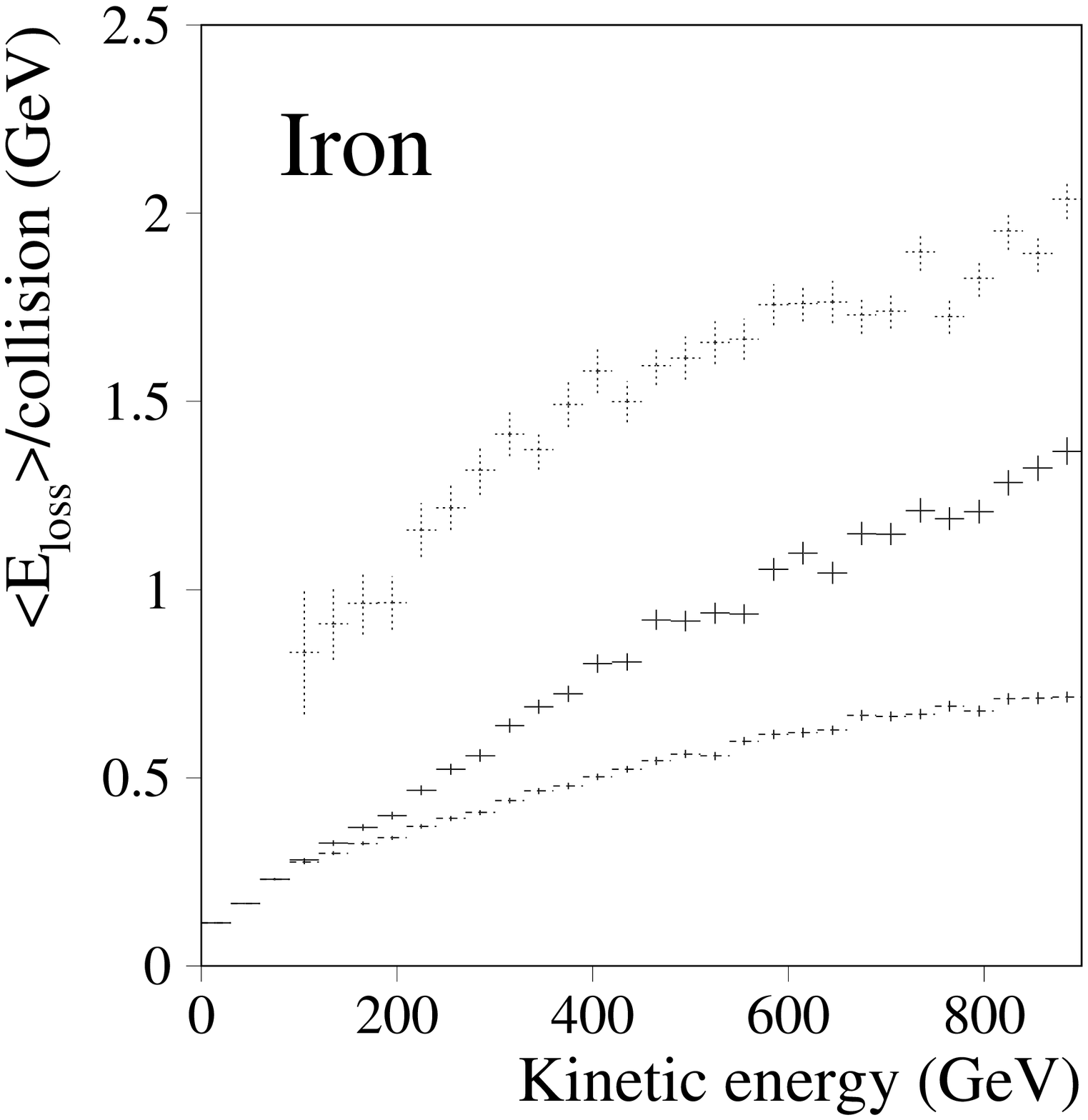,height=5.1cm,width=5.1cm}\hspace{0.4cm}\epsfig{file=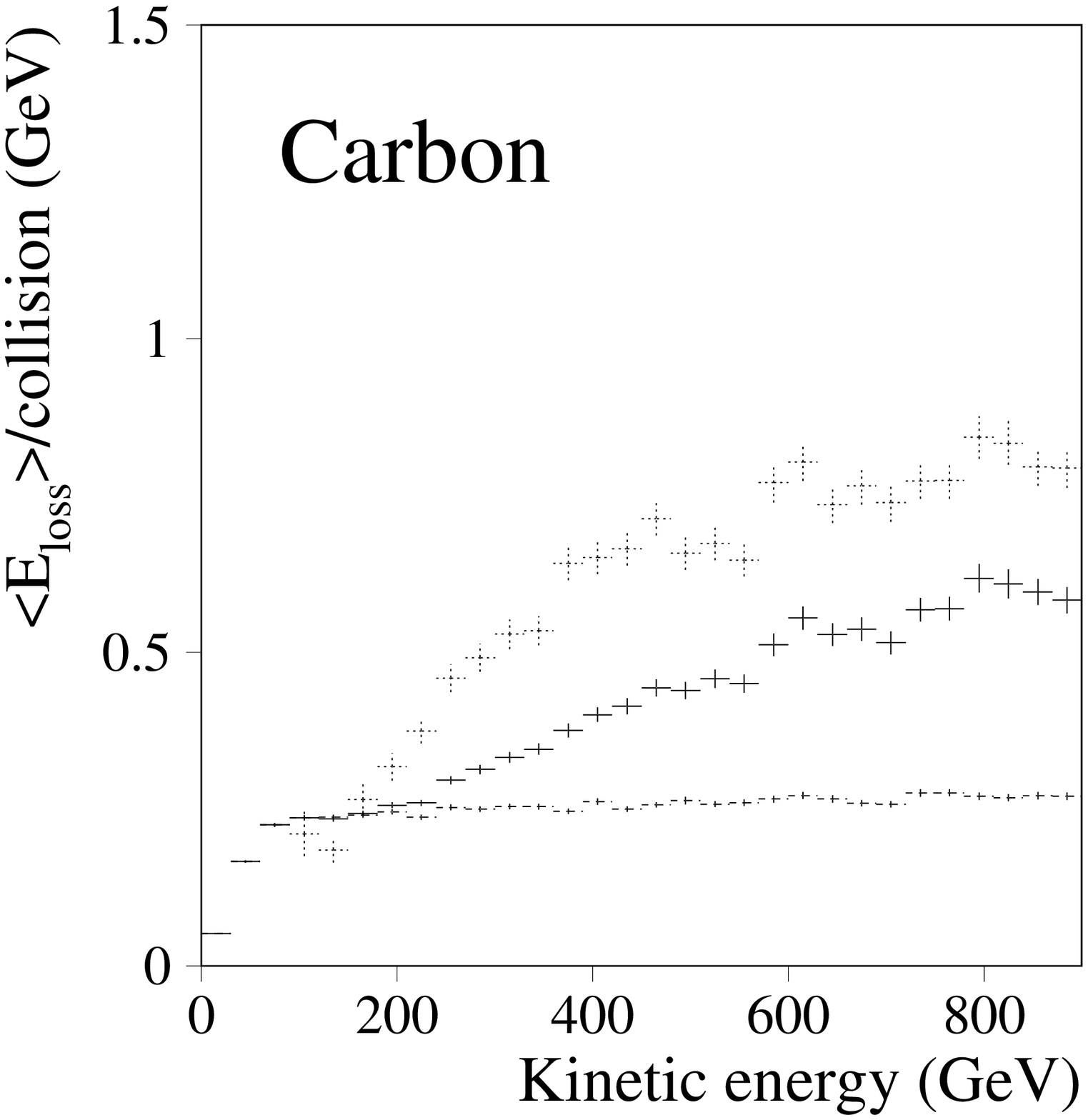,height=5.1cm,width=5.1cm}\hspace{0.4cm}\epsfig{file=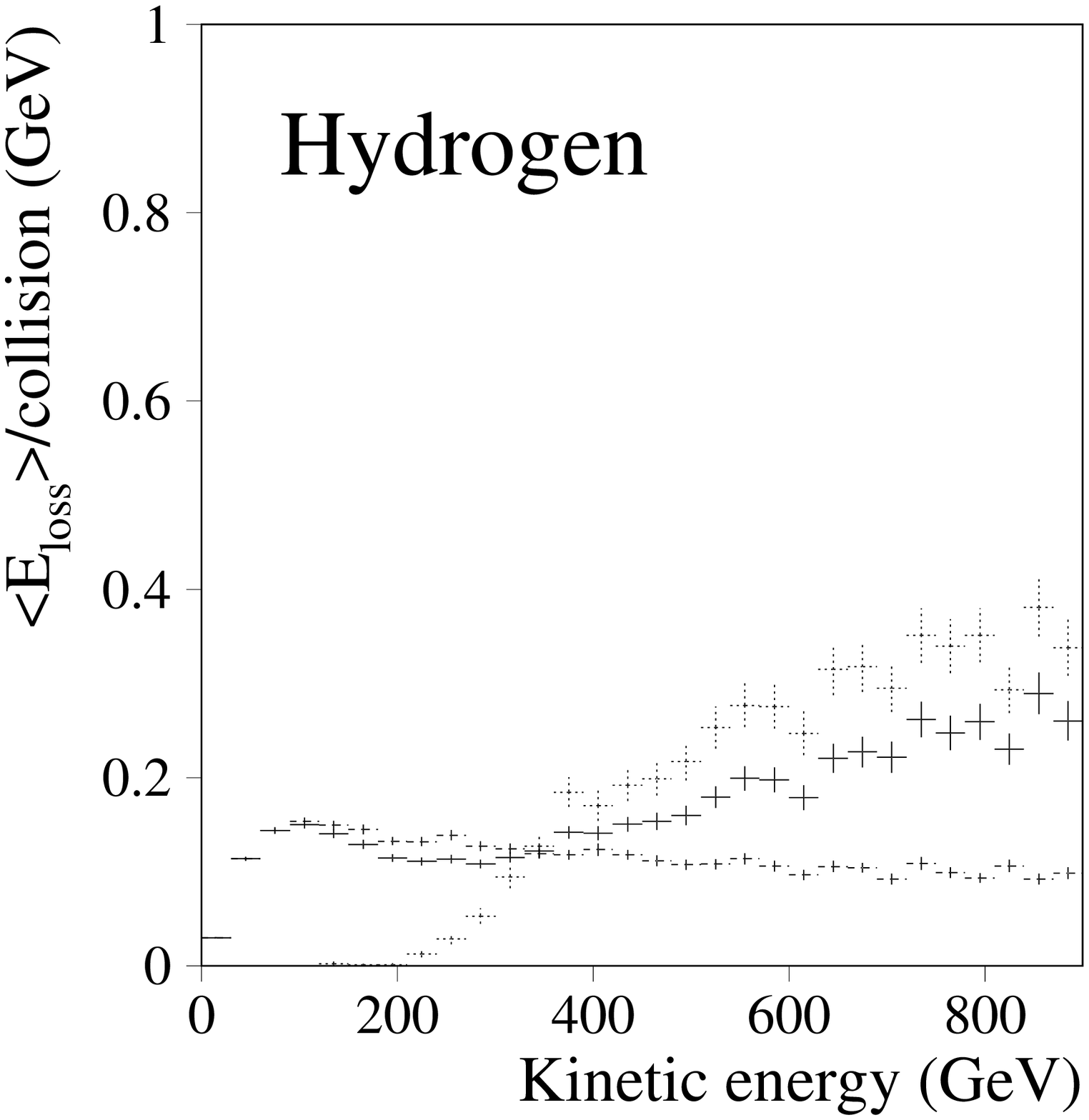,height=5.1cm,width=5.1cm}
\end{center}
\caption{Profile plots representing the mean value of the average energy loss per collision of an R--baryon for three different values of the R--hadron mass: 100 (three upper Figures), 300 (three middle figures) and 600 GeV/$c^2$ (three lower figures). The lines represent the mean value and its error. The lower and upper profiles in each plot correspond to the energy loss in a $2\rightarrow 2$ scattering and a $2\rightarrow 3$ scattering, respectively, while the middle profiles represent the average loss per collision.\label{fig:scat}}
\end{figure}

\subsection{Losses per collision}
To check whether non-trivial nuclear effects are simulated correctly, the losses per nuclear collision have been evaluated in iron, carbon and liquid hydrogen. In Fig.~\ref{fig:scat}, the losses per collision are displayed for $2\rightarrow 2$ and $2\rightarrow 3$ scatterings in the three materials for different values of the R--hadron mass. From the three figures obtained for e.g.~iron, it can be seen that the loss per collision decreases with increasing R--hadron mass for fixed kinetic energy. This is logically explained by the fact that the energy available for the production of new particles in a scattering becomes smaller. From these plots, we also note that the energy loss per collision is generally much larger for a $2\rightarrow 3$ processes than for $2\rightarrow 2$ processes, which follows from kinematics. If we compare different target materials, the loss per collision increases for heavy elements. This can be explained by the fact that the R--hadron is slow, and can scatter several times with the different nucleons inside one nucleus. For hydrogen, it should be noted that the energy loss for $2\rightarrow 3$ processes vanishes at small momenta, a feature which is not seen for heavier elements, where this effect is washed out by the nuclear cascade. This is due to a technical problem in the simulation package GHEISHA, which is entirely based on parameterizations for pions. Repairing this 'bug', and making the package optimal for heavy hadrons, would require a considerable rewrite of the code. This is beyond the scope of this work and, besides, 2$\rightarrow$2 scattering processes dominate completely at low energies. Energy losses for an R--hadron which converts from being a meson into baryon are slightly smaller when considered from the R--hadron perspective, but the total energy deposited in the detector is larger due to the larger kinetic energy released in the process. 
\begin{figure}[t!]
\begin{center}
\epsfig{file=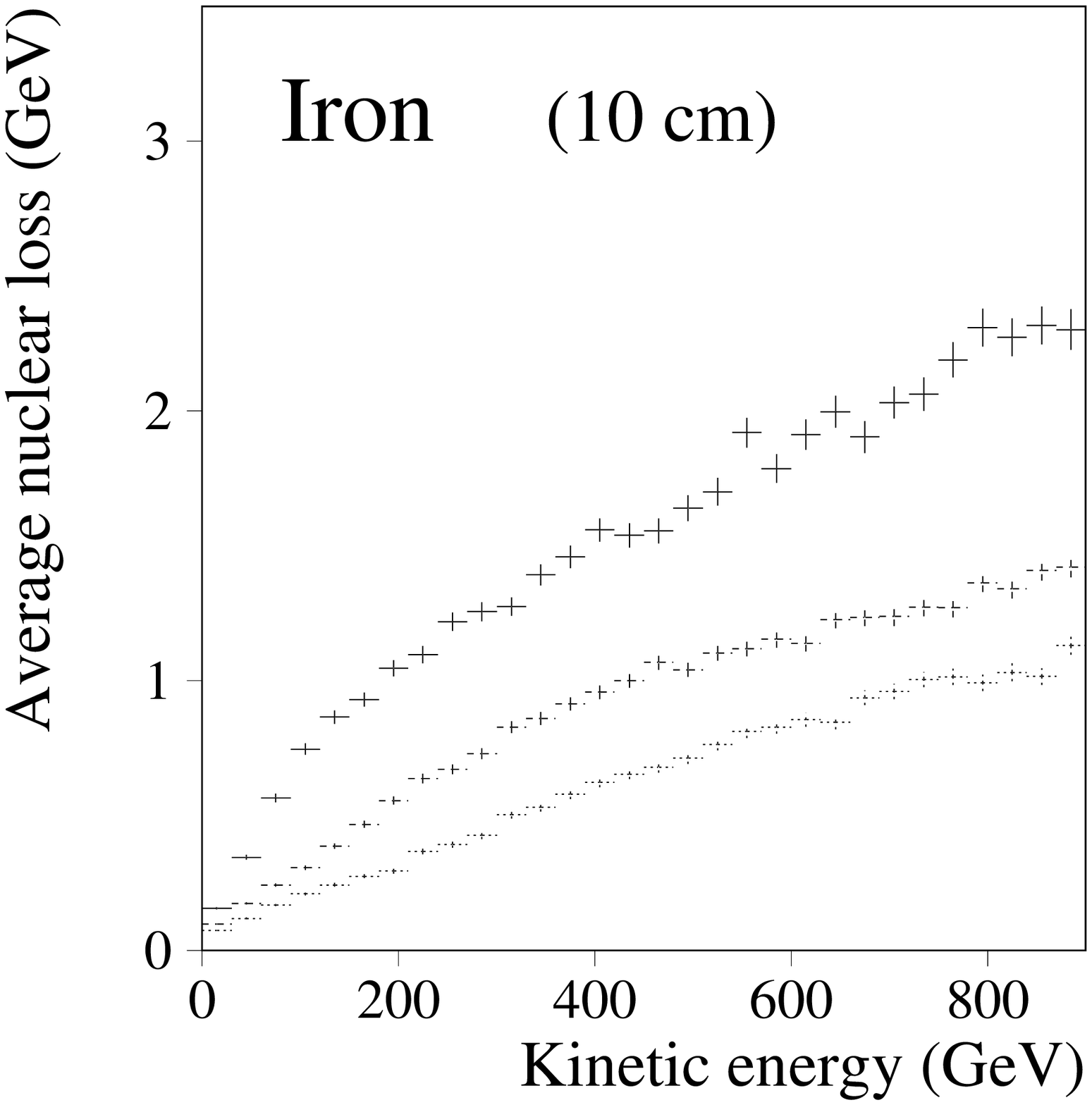,height=5.1cm,width=5.1cm}\hspace{0.4cm}\epsfig{file=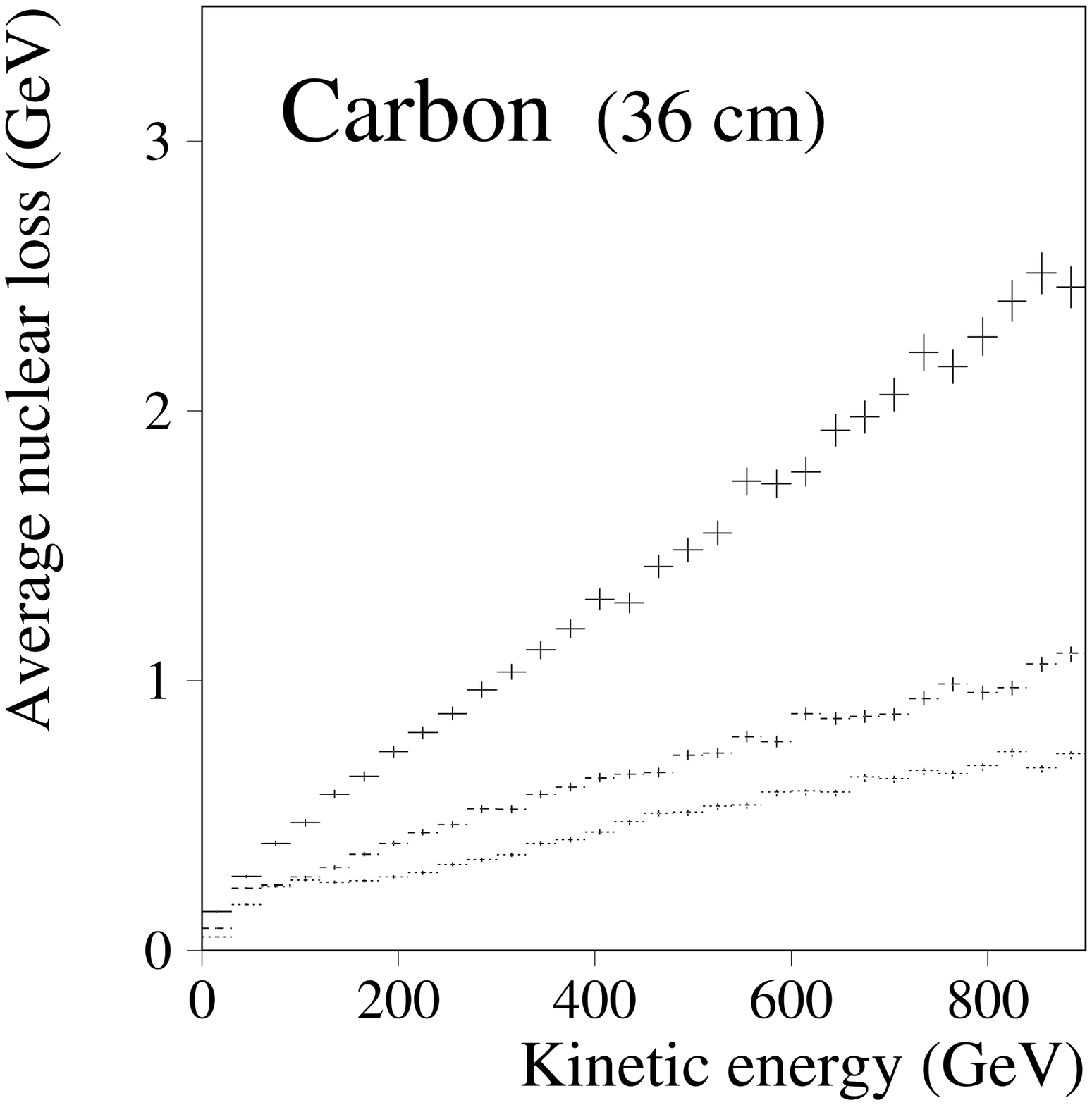,height=5.1cm,width=5.1cm}\hspace{0.4cm}\epsfig{file=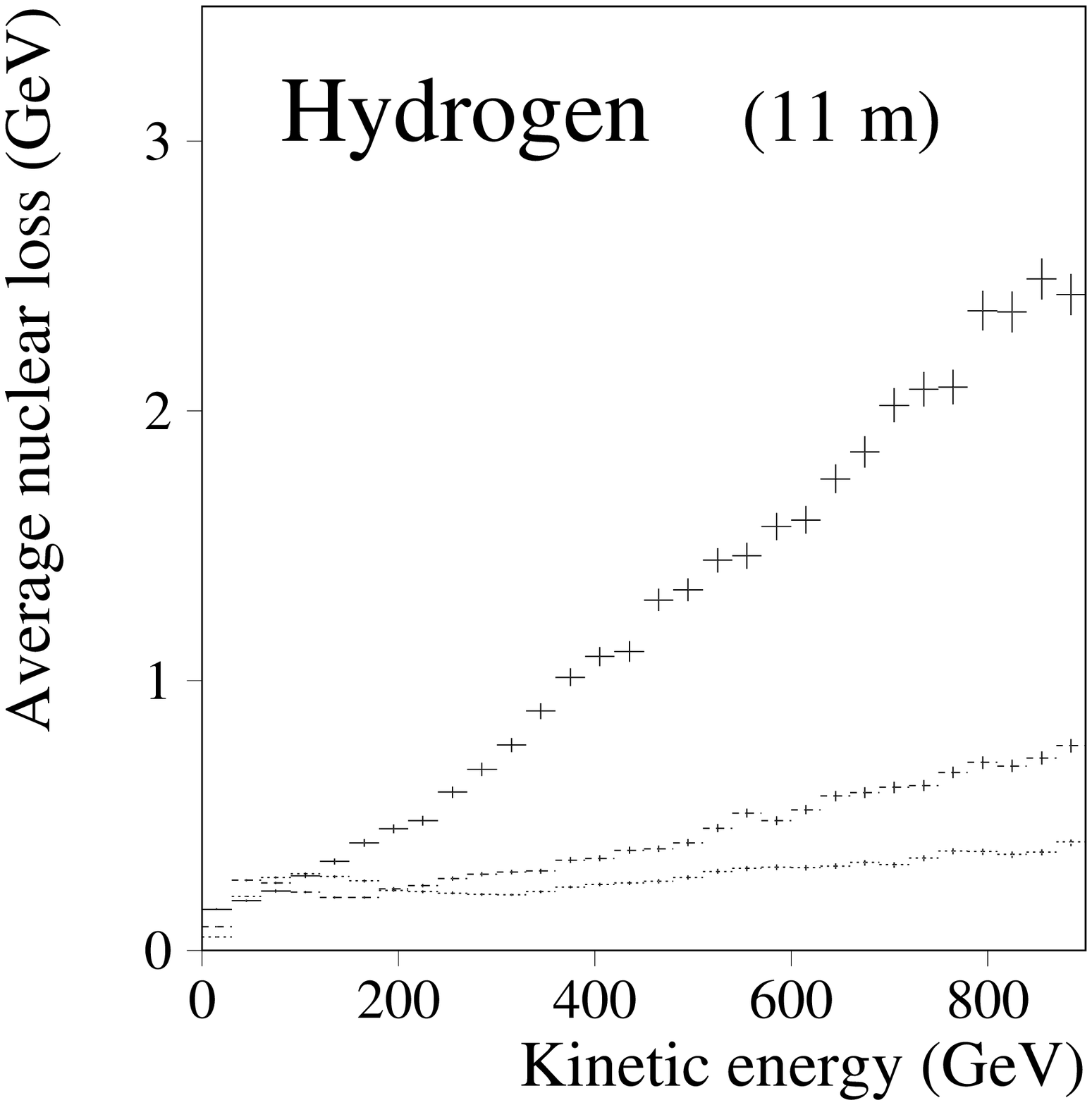,height=5.1cm,width=5.1cm}
\end{center}
\caption{Comparison of the total nuclear energy losses in 10 cm iron, in 36 cm carbon and in 11 m liquid hydrogen, corresponding to the same amount of nucleons, for an R--hadron mass of 100 GeV (upper curves), 300 GeV (middle curves) and 600 GeV (lower curves). \label{fig:10cm}}
\end{figure} 
\subsection{Nuclear energy losses in different materials}
We compare the energy loss to purely nuclear interactions (i.e.~no ionization losses) of an R--hadron in different materials, when the total amount of nucleons traversed is the same. Total nuclear losses of R--baryons in 10 cm iron, 36 cm carbon and 11 m liquid hydrogen are shown in Fig.~\ref{fig:10cm}. It turns out that approximately the same amount of total energy is lost. The average number of nuclear interactions is however not the same. As can be easily derived from the cross section on free nucleons, the interaction lengths in iron, carbon and  liquid hydrogen of an R--baryon are 14 cm, 32 cm and 610 cm, respectively. In 10 cm iron, 36 cm carbon, and  11 m liquid hydrogen ca 0.7, 1.1 and 1.8 collisions take place. However, the amount of energy lost in a collision increases with increasing atomic number (see Fig.~\ref{fig:scat}), and the overall result is that approximately the same amounts of energy are lost. 

\subsection{Total energy losses}
The total energy loss in 1 m iron, i.e.~the sum of nuclear energy losses and ionization losses, is determined for different mass values of the R--hadron, and results are displayed in Fig.~\ref{fig:1m}. The ionization losses for a singly charged R--hadron traversing 1 m iron without suffering any nuclear interactions are shown as well. Nuclear losses are seen to be completely dominating at high energies.

\subsection{Nuclear energy losses versus ionization losses}
\begin{figure}[t!]
\begin{center}
\epsfig{file=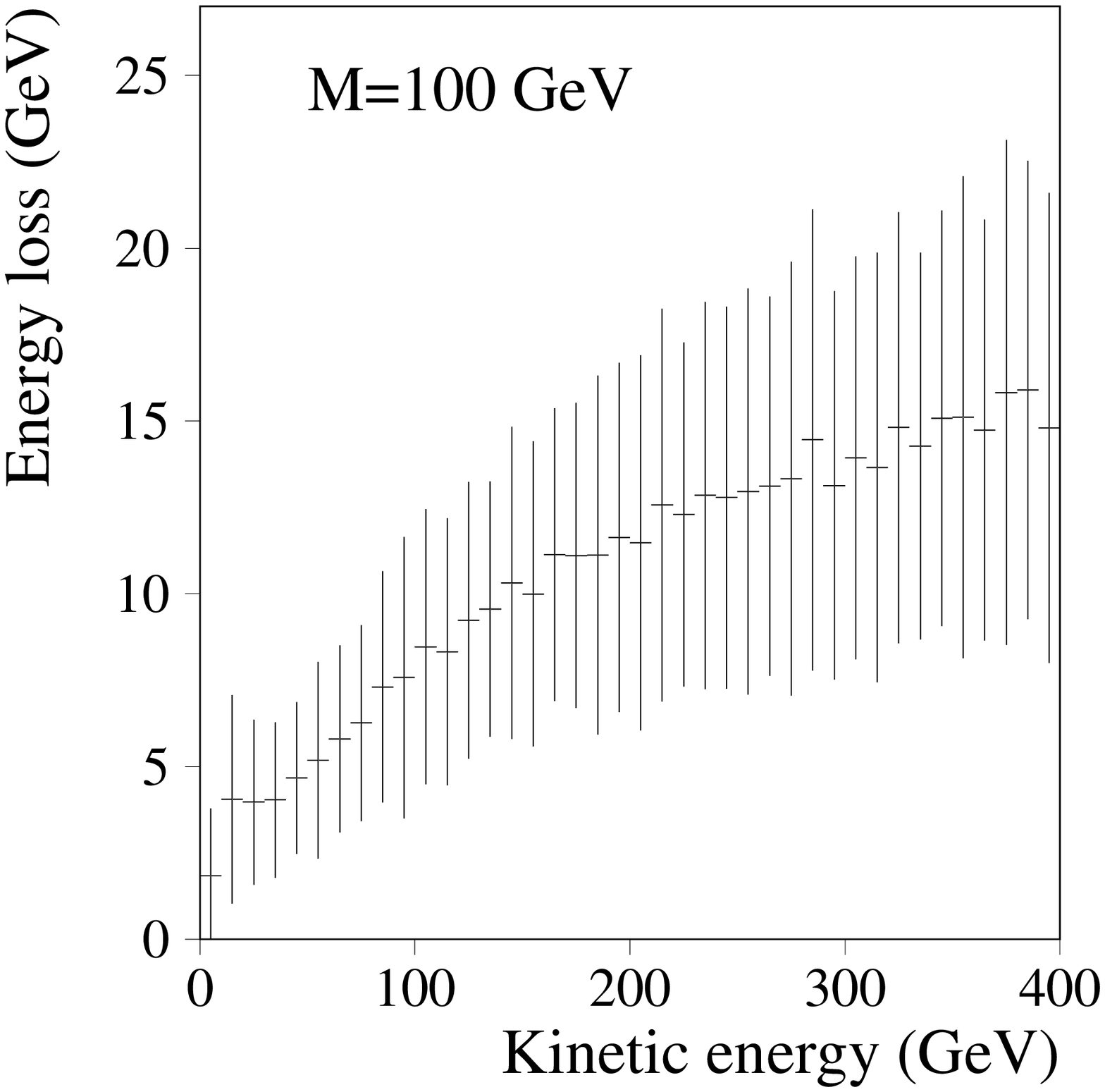,height=5.1cm,width=5.1cm}\hspace{0.3cm}\epsfig{file=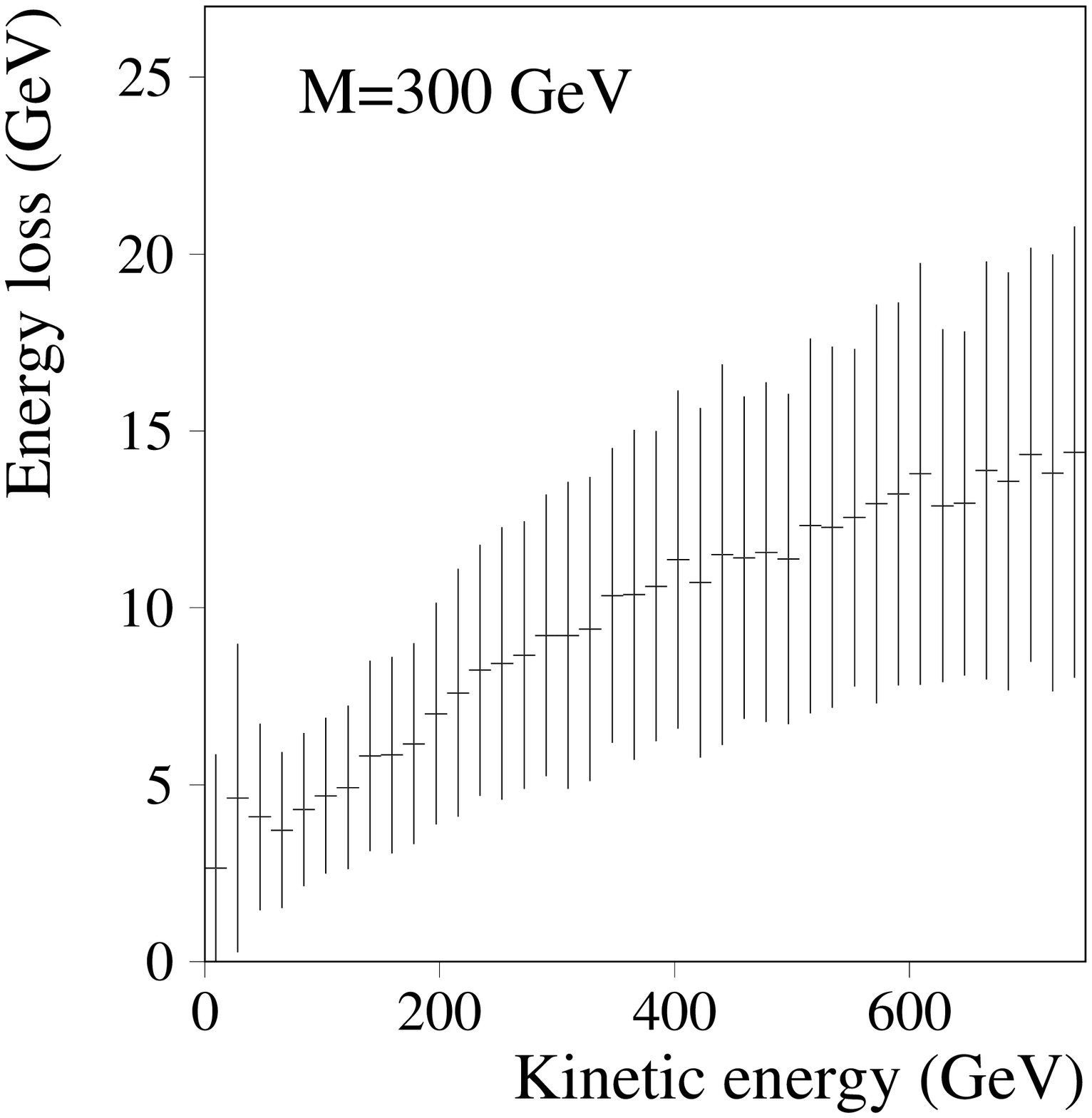,height=5.1cm,width=5.1cm}\hspace{0.3cm}\epsfig{file=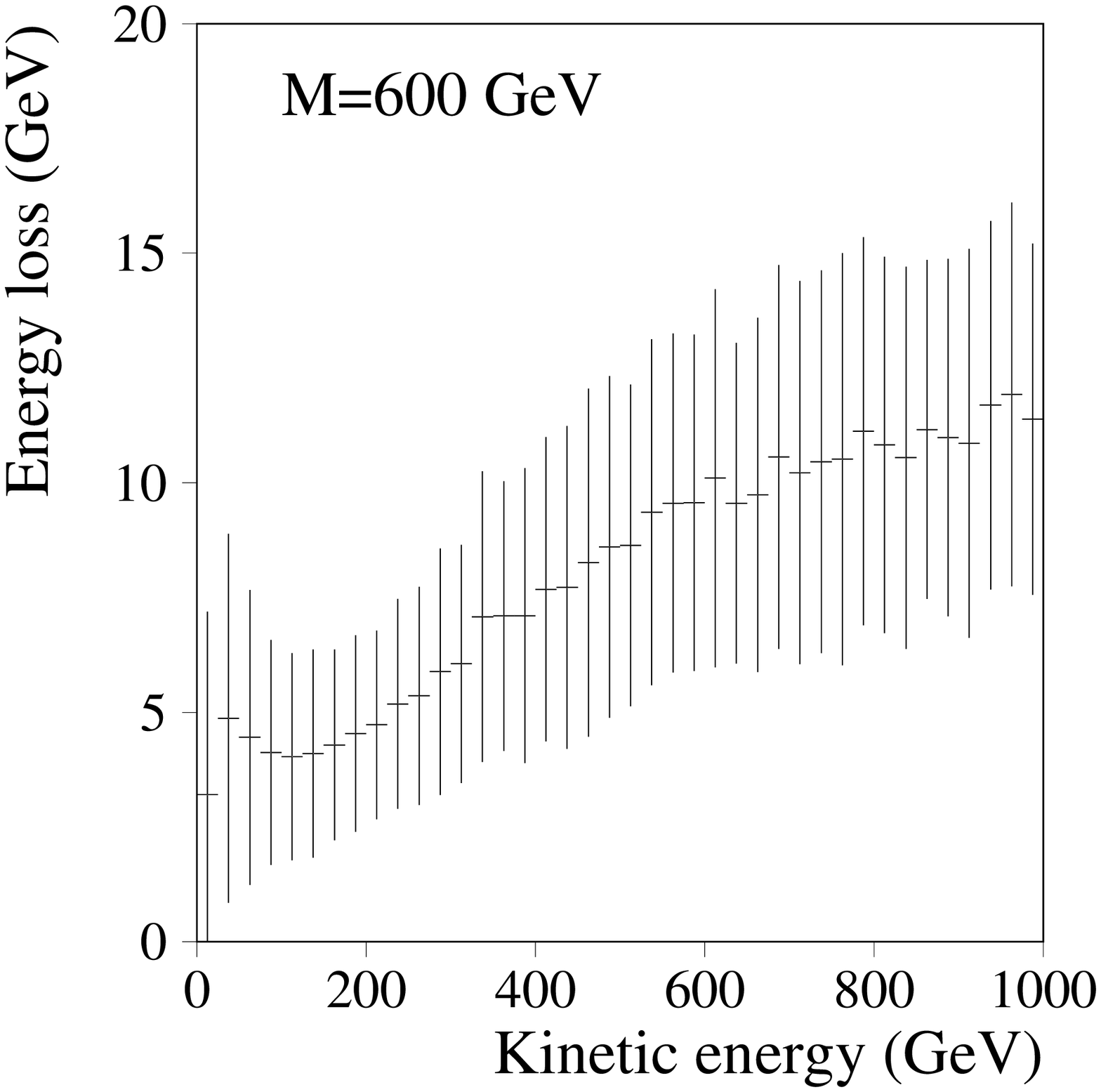,height=5.1cm,width=5.1cm} \\
\epsfig{file=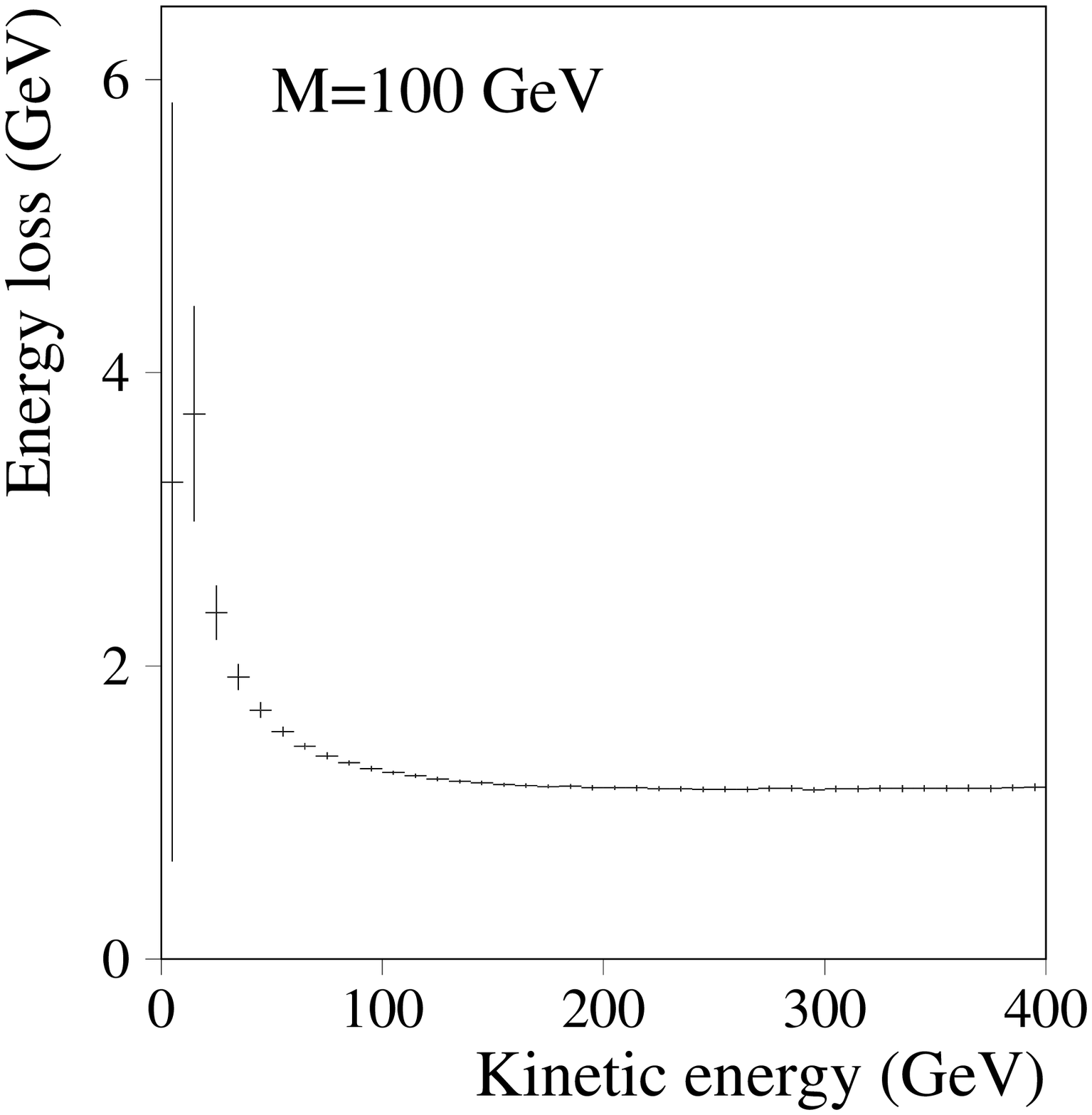,height=5.1cm,width=5.1cm}\hspace{0.3cm}\epsfig{file=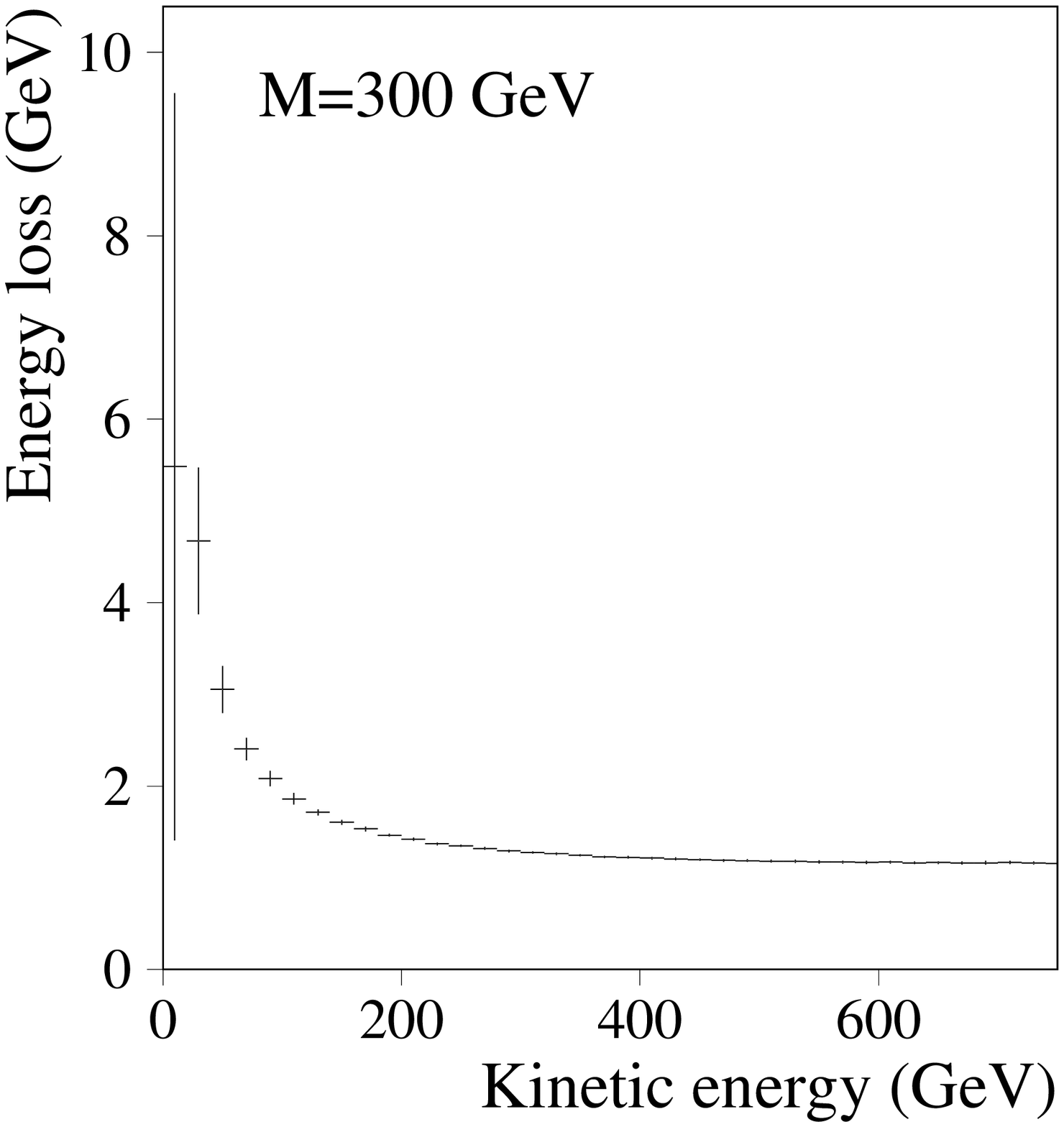,height=5.1cm,width=5.1cm}\hspace{0.3cm}\epsfig{file=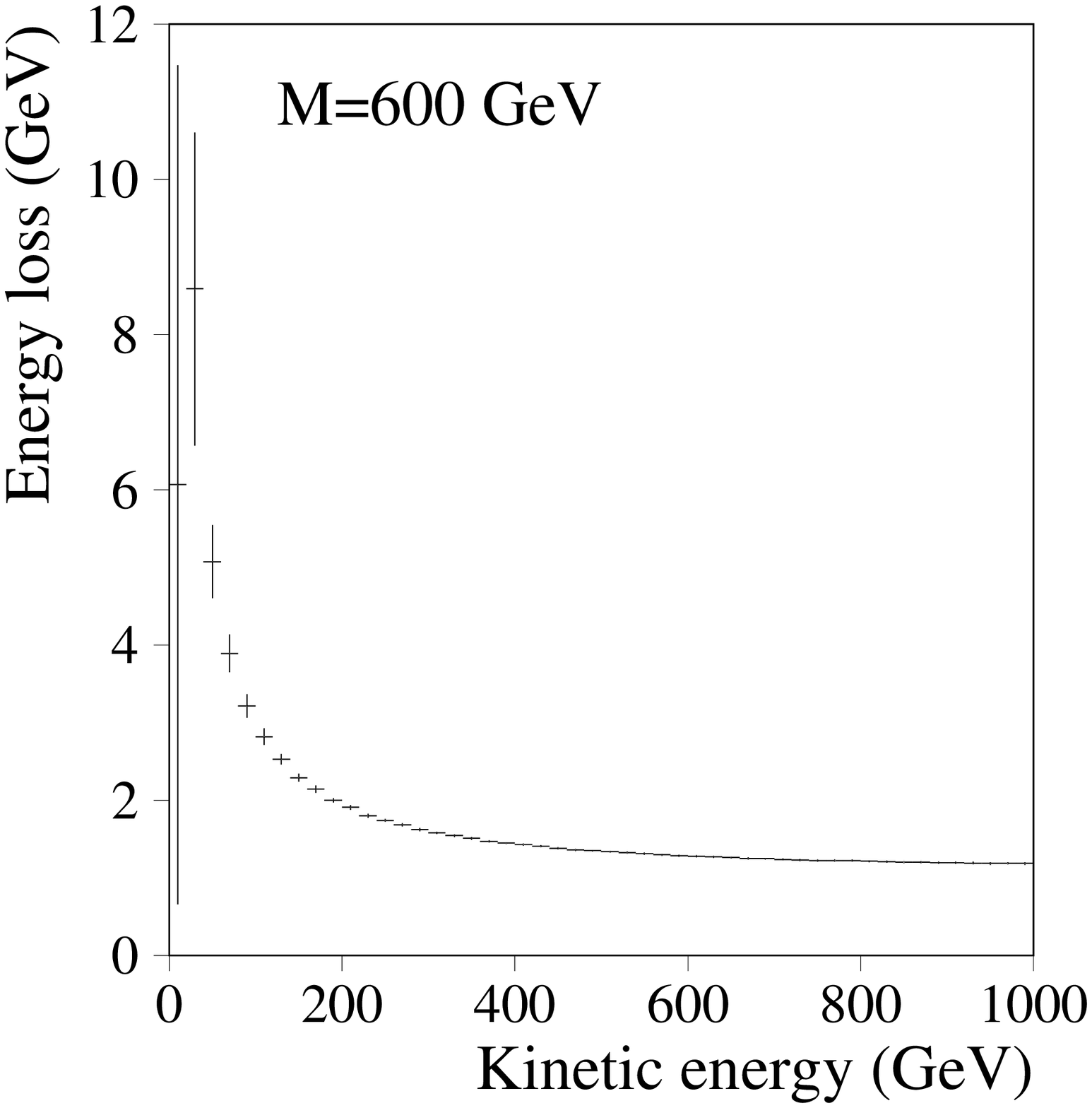,height=5.1cm,width=5.1cm}
\end{center}
\caption{Profile histograms representing the average R--hadron energy loss in one meter iron and the spread on this average as a function of its kinetic energy for the total energy loss (upper figures) and pure ionization losses for a singly charged R--hadron not undergoing nuclear interactions (lower figures).\label{fig:1m}} 
\end{figure}

\begin{figure}[t!]
\begin{center}
\epsfig{file=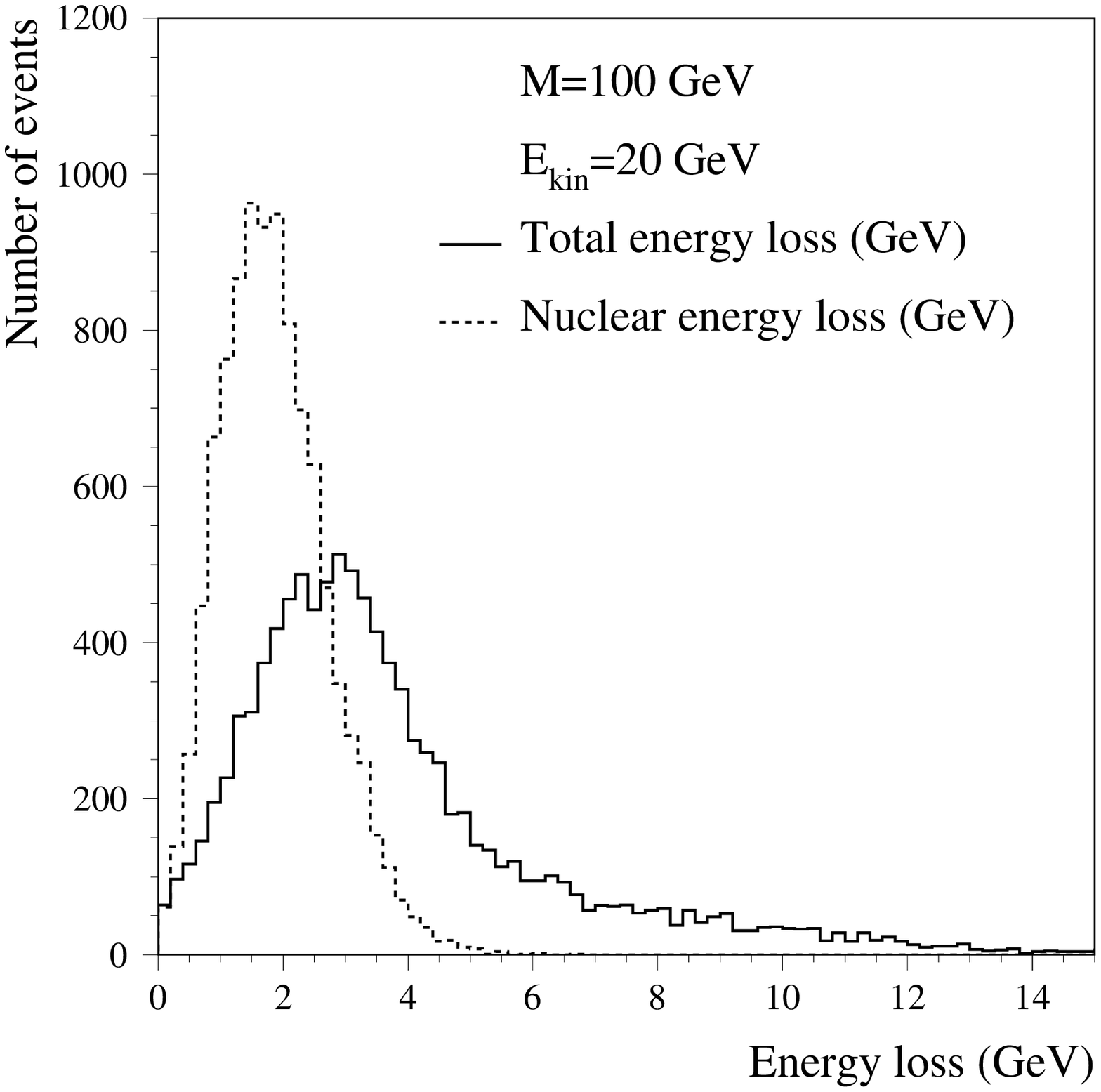,height=7.7cm,width=7.7cm}\hspace{0.5cm}\epsfig{file=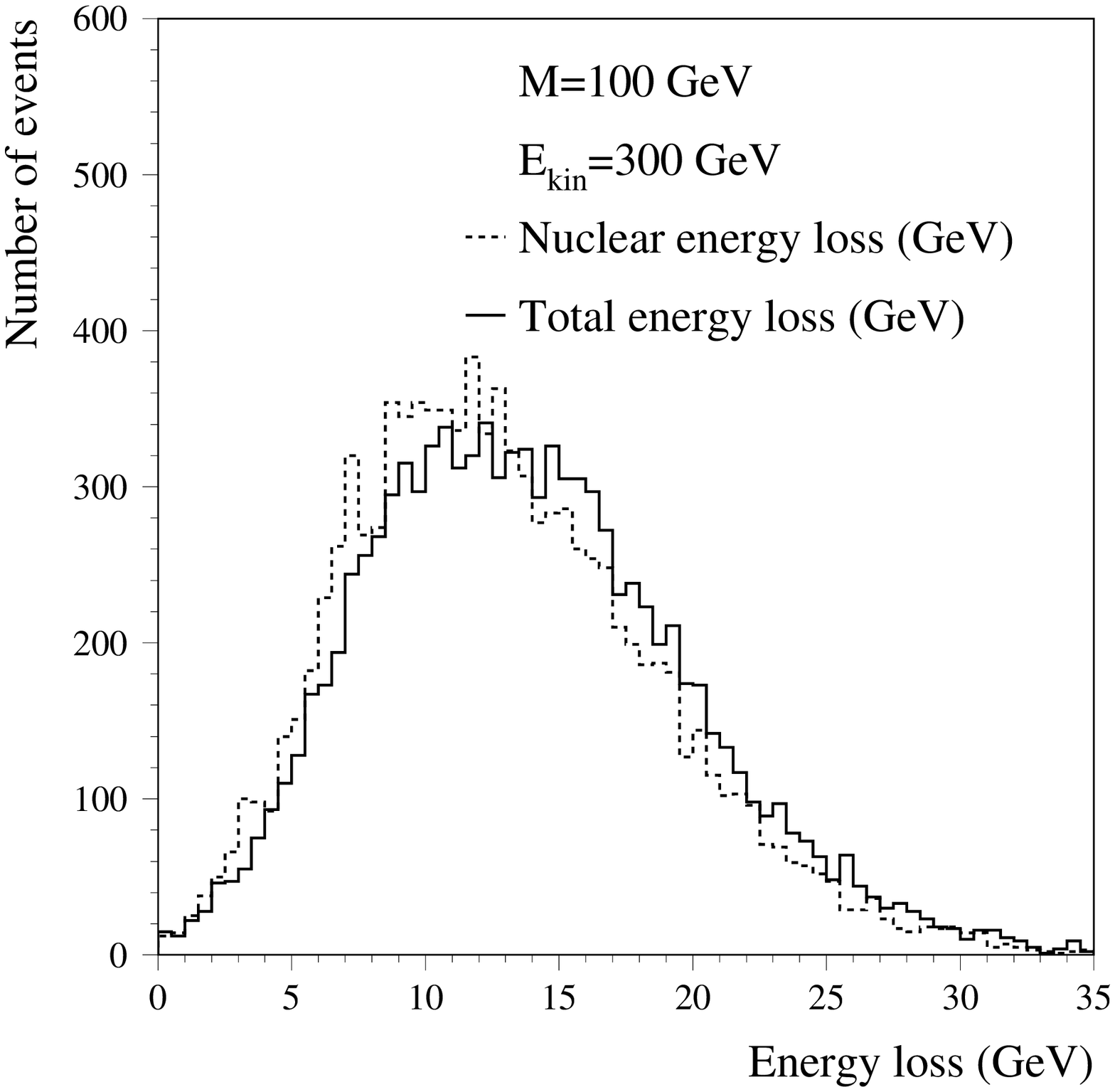,height=7.5cm,width=7.7cm}\\
(a) \hspace{5cm}(b)
\end{center}
\caption{A comparison of total energy loss and the nuclear energy loss alone in 1 meter iron for an R--hadron with mass 100 GeV for a kinetic energy of 20 GeV (a) and of 300 GeV (b). The mean values of the distributions shown in Fig.~(a) are 3.8 and 1.9 respectively, while those in Fig.~(b) are 13.8 and 12.9 respectively.  \label{fig:direct}}
\end{figure}

A direct comparison between the energy losses of an R--hadron suffering only nuclear collisions, and those of an R--hadron suffering ionization losses  as well as nuclear collisions, is shown in Fig.~\ref{fig:direct}.  At a kinetic energy of 300 GeV, ionization losses are responsible for a shift of the mean value of the energy loss of 0.9 GeV, slightly less than the value displayed in the ionization figure for $M=100$ GeV in Fig.~\ref{fig:1m}. This is obviously due to the fact that the R--hadron may change charge and hence may be neutral during part of the trajectory. At a kinetic energy of 20 GeV, the shift is larger, 1.9 GeV for this particular R--hadron mass. The conclusion is that for low--energetic heavy hadrons, ionization losses are comparable to nuclear losses, while at high kinetic energies, nuclear losses are completely dominating, as expected.

\subsection{Dependence of energy losses on the phase space function}
To evaluate the dependence on the parameter $Q_0$, appearing in the phase space function $F(Q)$, it is varied from 0.6 GeV/$c^2$ to 1.7 GeV/$c^2$. From Fig.~\ref{fig:phasespace}, we note that the available phase space for $2\rightarrow 3$ scattering decreases slightly with increasing value of $Q_0$, and therefore the energy loss is expected to decrease for increasing $Q_0$.  For two values of the R--hadron mass, the total energy loss in 1 meter iron is studied for different values of $Q_0$, and results are shown in Table~\ref{tab:epsilon}. The dependence on $Q_0$ of the total energy loss is small, and indeed losses are smaller for increasing $Q_0$-value.

\subsection{Penetration depth}
\begin{figure}[h!]
\begin{center}
\epsfig{file=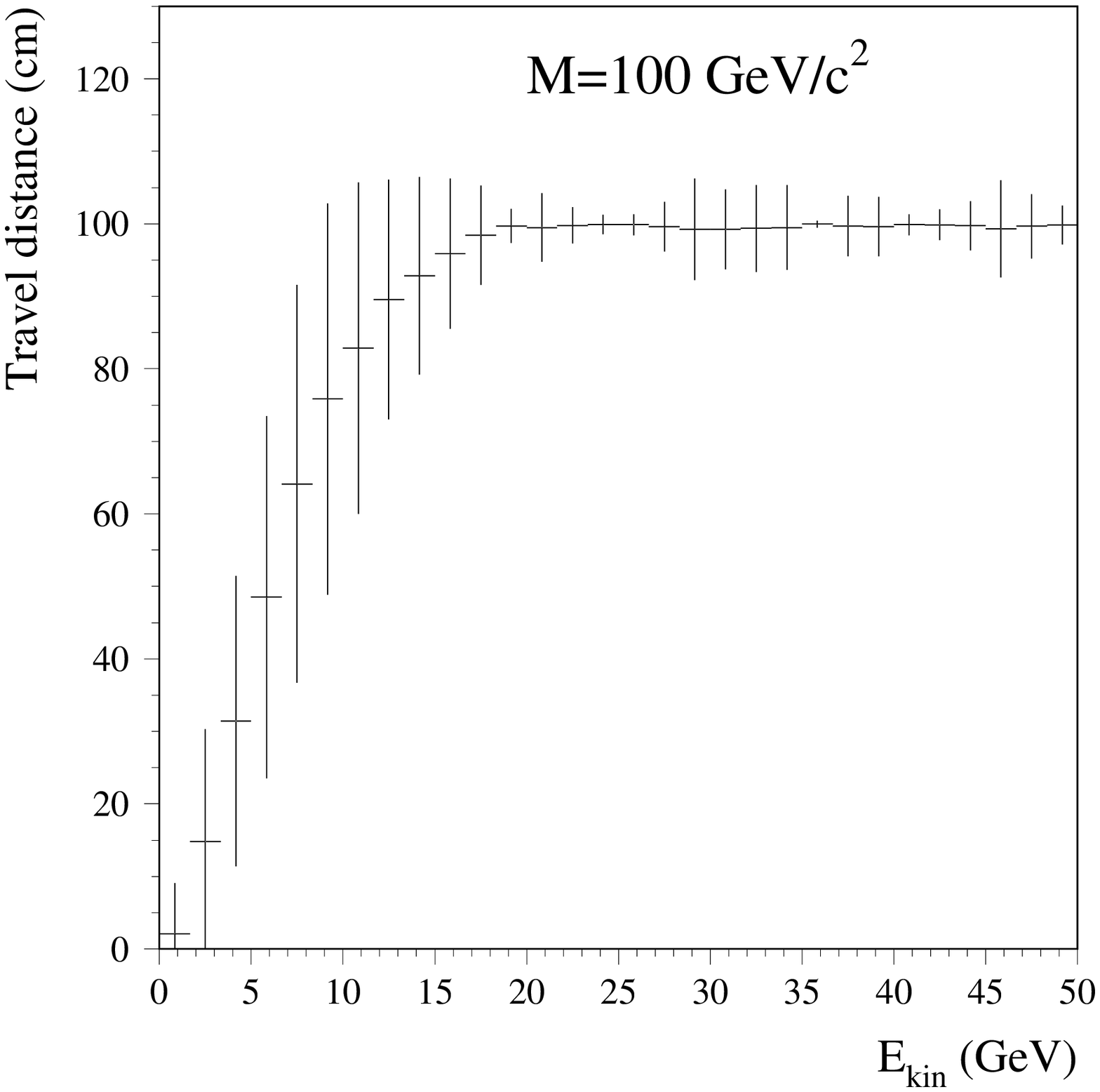,height=5cm,width=5cm}\hspace{0.5cm}\epsfig{file=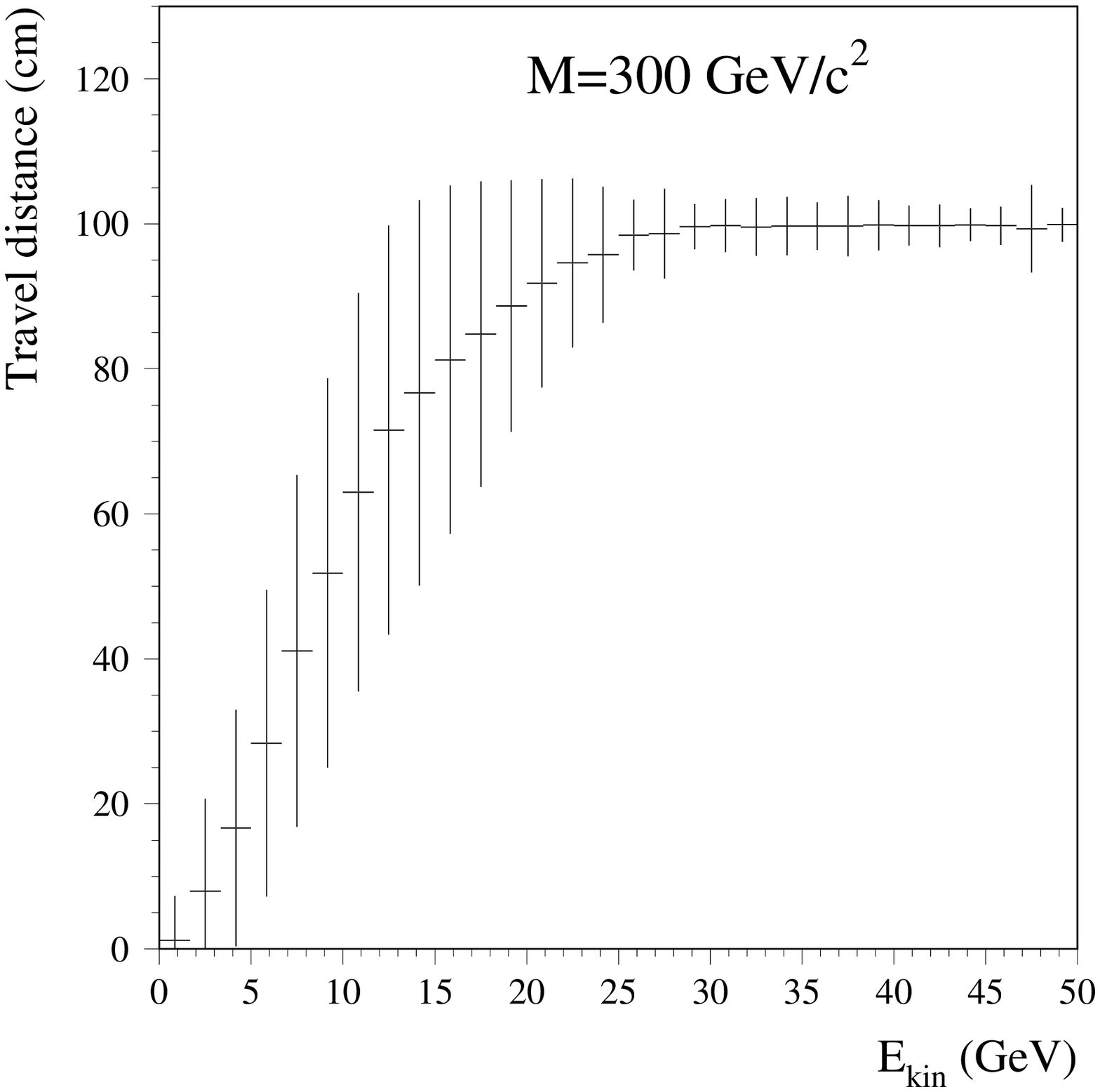,height=5cm,width=5cm}\hspace{0.5cm}\epsfig{file=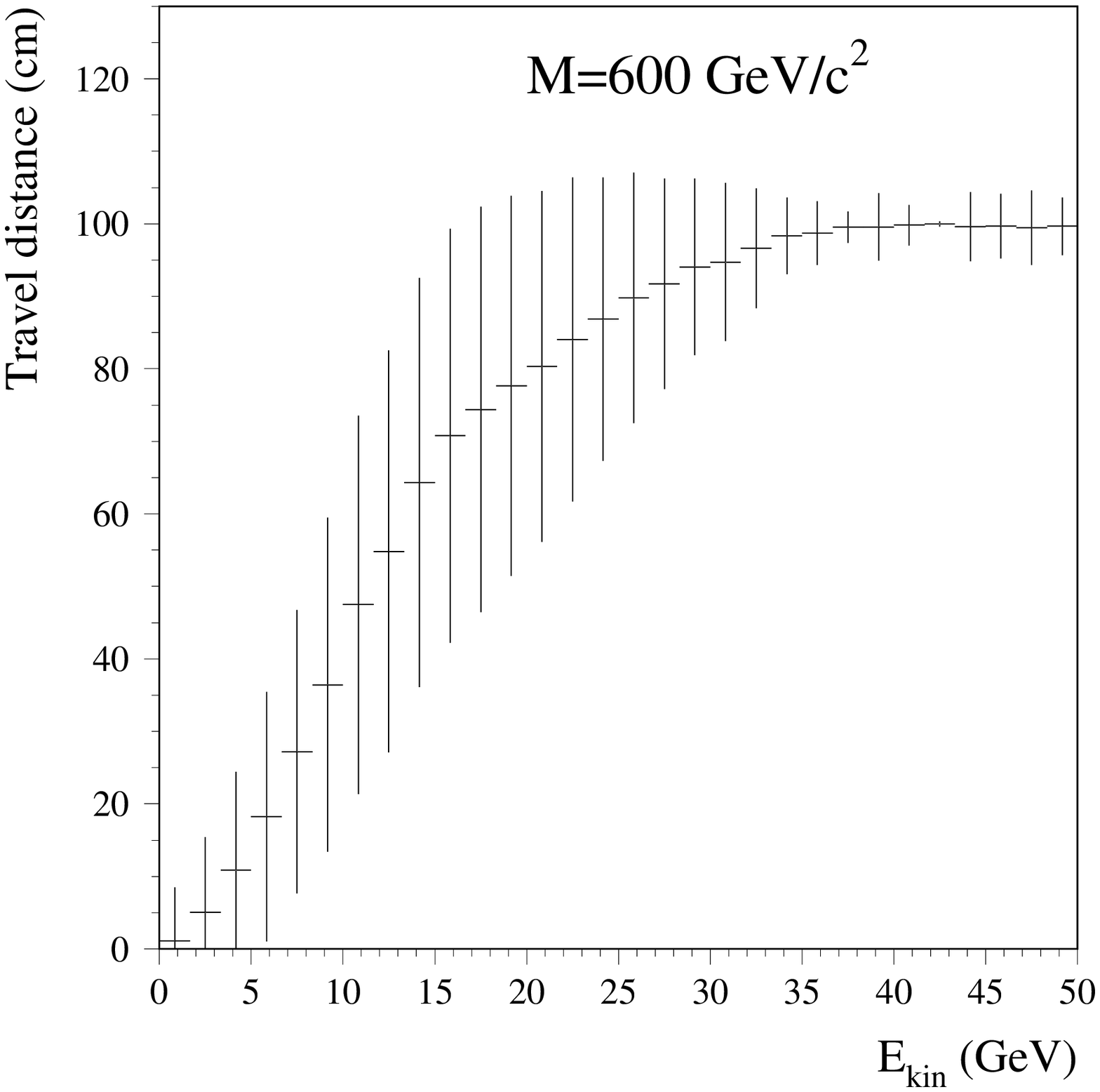,height=5cm,width=5cm}
\end{center}
\caption{Comparison of traversed path of a singly charged R--baryon for different values of the R--hadron mass. The vertical lines represent the spread on the mean value.\label{fig:iets}}
\end{figure}

In order to estimate the penetration depth of R--hadrons, a study is done to the traversed path of an R--baryon in 1 m iron. The energies of R--hadrons which do not get stopped in the calorimeters is relevant, since those hadrons reach the outer parts of a typical detector; often the muon chambers. In Fig.~\ref{fig:iets}, the traversed length in iron is shown for different R--hadron masses, with a maximum length of 1 m. A rapid saturation takes place, i.e.~above a rather moderate value R--hadrons will punch through a typical calorimeter.

\section{Signatures of heavy hadrons in ATLAS}\label{atlasref}
\begin{figure}[t!]
\begin{center}
\epsfig{file=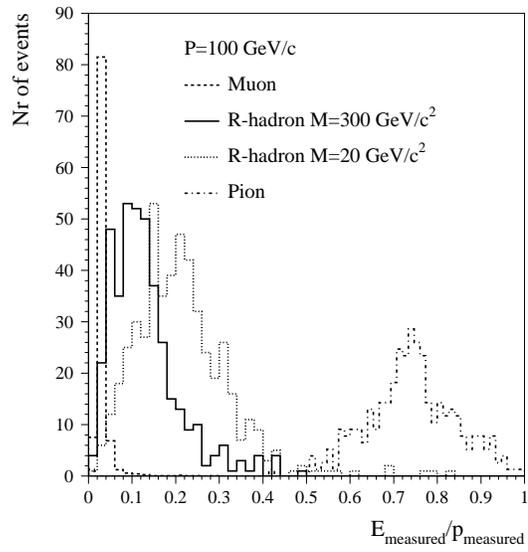,height=8cm,width=8cm}
\end{center}
\caption{The ratio $E/p$ of the deposited energy in the ATLAS calorimeters and the momentum, as measured in the ATLAS tracking system for particles generated with momentum $p$ = 100 GeV/$c^2$. The scale on the y-axis is arbitrary. \label{fig:ep}}
\end{figure}

As an example of R--hadron signatures in a typical future detector, we consider the ATLAS detector. Concerning the selection of events containing R-hadrons, appropriate triggers would be the muon trigger, in case a charged R--hadron reaches them, or by the missing energy trigger, if not two R--hadrons are produced in a back-to-back topology in the transverse plane. Signatures would include the following.
\begin{itemize}
\item{Missing energy, the amount depending on the topology of the events.}
\item{A larger amount of ionization in the Transition Radiation Tracker, manifested by a large amount of high threshold hits, if the R-hadron moves slowly. This could mimic the transition radiation hits for particles with very large $\beta$ values. }
\item{The $E/p$ ratio, which is the amount of energy deposited in the calorimeters, divided by the momentum of the track, measured in the tracking system. This ratio is considerably smaller than that for light hadrons like pions, but larger than that for a muon, as is illustrated Fig.~\ref{fig:ep}.}
\item{A characteristic longitudinal shower profile in the different components in the ATLAS calorimeter, which would differ from that of pions, as well as from that of muons.}
\item{A characteristic transverse shower profile in the calorimeters. The low--energetic interactions result in a narrower shower than e.g. a shower caused by a high--energetic pion. }
\item{A large time-of-flight in the muon chambers.}
\end{itemize}
A detailed study of all the above aspects, as well as a detailed trigger efficiency study for the relevant triggers in the ATLAS experiment, will be discussed in a future publication. 

\section{Conclusion}\label{con}
In this paper, interactions of heavy hadrons in matter have been described, and their interactions have been simulated inside the GEANT 3 framework. Several approximations are made in the model with respect to cross sections, quark content, and interaction processes. The model presented here, as well as the simulation, should not be viewed as a final description, but provides a convenient platform on which to build in the future.

\section*{Acknowledgments}
Special thanks to Torbj{\"o}rn Sj{\"o}strand for many valuable discussions, and for his useful comments and ideas for this paper. Furthermore, I thank Pavel Nevski for his help and advice with the GEANT 3 simulation. I also would like to acknowledge useful discussions with Peter Hansen about experimental issues. Finally, thanks to Andrew D. Jackson for the discussion about nucleon scattering cross sections.

\begin{table}[b!]
\begin{center}
\begin{tabular}{|c|c|c|c|}
\hline
 Mass (GeV/$c^2$) &Kinetic energy (GeV) & $Q_0$ (GeV/$c^2$)  & $<E_{loss}>$ in 1 m iron  (GeV) \\
\hline
\hline
 & & 0.6 & 3.8\\
100 & 20& 1.1 & 3.8  \\
 & & 1.7 & 3.8\\
\hline
  & & 0.6 & 14.3\\
100 & 300 & 1.1  & 13.8 \\
 & & 1.7 & 13.3\\
\hline
  & & 0.6 & 20.5\\
100 & 500 & 1.1  & 20.2 \\
 & & 1.7 & 19.9\\
\hline
& & 0.6 & 6.5\\
600 & 20 & 1.1  & 6.5\\
 & & 1.7 & 6.5\\
\hline 
& & 0.6 & 6.8\\
600 & 300 & 1.1 & 6.0\\
 & & 1.7 & 5.6\\
\hline
& & 0.6 & 12.1\\
600 & 500 & 1.1 & 10.6\\
 & & 1.7 & 9.6 \\
\hline
\end{tabular}
\end{center}
\caption{The mean value of the energy loss in 1 m iron for different values of $Q_0$. The default value is $Q_0=1.1$ GeV/$c^2$ \label{tab:epsilon}}
\end{table}

\end{document}